\documentclass[twocolmns]{aastex631}

\begin{document}

\title{ Relative astrometry in an annular field}

\author{M. Gai}
\affiliation{INAF - Osservatorio Astrofisico di Torino --
	I-10025 Pino Torinese (TO), Italy}

\author{A. Vecchiato}
\affiliation{INAF - Osservatorio Astrofisico di Torino --
	I-10025 Pino Torinese (TO), Italy}

\author{A. Riva}
\affiliation{INAF - Osservatorio Astrofisico di Torino --
	I-10025 Pino Torinese (TO), Italy}

\author{A. G. Butkevich}
\affiliation{INAF - Osservatorio Astrofisico di Torino --
	I-10025 Pino Torinese (TO), Italy}
\affiliation{Pulkovo Observatory, Russian Academy of Sciences -- 
	Saint Petersburg 196140, Russia}

\author{D. Busonero}
\affiliation{INAF - Osservatorio Astrofisico di Torino --
	I-10025 Pino Torinese (TO), Italy}

\author{Z. Qi}
\affiliation{Shanghai Astronomical Observatory, Chinese Academy of Sciences -- 
	Shanghai 200030, China}

\author{M. G. Lattanzi}
\affiliation{INAF - Osservatorio Astrofisico di Torino --
	I-10025 Pino Torinese (TO), Italy}

\date{October 2021}

\begin{abstract}
{{\it Background.} 
Relative astrometry at or below the micro-arcsec level with a $1\,m$ class 
space telescope has been repeatedly proposed 
as a tool for exo-planet detection and characterization, as well as for several 
topics at the forefront of Astrophysics and Fundamental Physics. }
{{\it Aim.} 
This paper investigates the potential benefits of an instrument concept based on 
an annular field of view, as compared to a traditional focal plane imaging a contiguous 
area close to the telescope optical axis. }
{{\it Method.} 
Basic aspects of relative astrometry are reviewed as a function of the distribution 
on the sky of reference stars brighter than $G = 12\,mag$ (from Gaia EDR3). 
Statistics of field stars for targets down to $G = 8\,mag$ is evaluated by 
analysis and simulation. }
{{\it Results.} 
Observation efficiency benefits from prior knowledge on individual targets, 
since source model is improved with few measurements. 
Dedicated observations (10-20 hours) can constrain the orbital inclination of 
exoplanets to a few degrees. 
Observing strategy can be tailored to include a sample of stars, 
materialising the reference frame, sufficiently large to average down the 
residual catalogue errors to the desired micro-arcsec level. 
For most targets, the annular field provides typically more reference stars, 
by a factor four to seven in our case, than the conventional field. 
The brightest reference stars for each target are up to $2\,mag$ brighter. 
}
{{\it Conclusions.} 
The proposed annular field telescope concept improves on observation flexibility 
and/or astrometric performance with respect to conventional designs. 
It appears therefore as an appealing contribution to optimization of future 
relative astrometry missions. }
\end{abstract}


\section{ Introduction }
\label{Sec:intro}
Astrometry from space bears the promise of achieving substantially photon-  
and diffraction-limited precision to measurement of celestial object positions 
(and related quantities: motion, parallax, separation, ...). 
Moreover, such precision may be achieved on either narrow or large angular separation 
among targets, providing the tools to achieve full-sky (global or absolute) 
measurements. 
The most advanced large scale implementation of such ideas is currently 
Gaia \citep{GaiaDR1Mission}, which has recently delivered its Early Data Release 3 (EDR3) 
\citep{GaiaEDR3Summary21}, with remarkable astrometric quality 
\citep{GaiaEDR3Astrometry21}, expected to further improve in future releases.  

However, as any scientific progress is a stepping stone toward the next challenge, 
a number of other space missions have been, and will be, proposed to further 
advance our capability of understanding the Universe by better measurements. 
Some of them are briefly recalled below. 

Photon-limited precision at the micro-arcsec (hereafter, $\mu as$) level, 
or better, translates into comparable accuracy only at the expense of 
significant efforts in terms of instrument design and calibration efforts, 
which also affects operation. 
Calibration of the optical response is obviously alleviated in case of a 
nearly ideal instrument, with small deviation from the diffraction limited 
Point Spread Function (PSF). 
Advanced metrology concepts, improving on the knowledge of instrument response, 
have also been investigated \citep{ZhaiShao2011}. 

The science case for high precision, narrow angle astrometry has been presented 
e.g. in the context of the proposed space mission Theia 
\citep{Malbet+2021}. 
Exo-planetary science missions based on astrometry have also been investigated 
in other approaches \citep{Shao19,Bendek18,TOLIMAN18}. 
Also observatory class missions like the Roman Space Telescope 
\citep{Croft21} bear the potential for impressive results, in that case 
with lower angular precision, but on much fainter extra-galactic 
objects. 
We retain as reference most of the goals of the above concepts, 
but we introduce a significant modification in the instrument and operation 
implementation concept, which we expect to be beneficial with respect 
to calibration and control of systematic errors related to variation of 
the telescope optical response. 

The recently proposed idea \citep{RAFTER_SPIE_20} of a Ring Astrometric Field 
Telescope for Exoplanets and Relativity (RAFTER), reviewed in 
Sec.\,\ref{Sec:rafter}, is 
characterised by a field of view, and related detection system, deployed 
over a circular strip centered around the projection of the optical axis. 
Also, circular symmetry is preserved throughout the optical system, thus 
ensuring circular symmetry of the instrument response, as described in 
Sec.\,\ref{Sec:instrument}.  

\begin{figure}[t!]
	\centering 
	\begin{minipage}[c]{0.47\textwidth}
\includegraphics[width=\hsize]{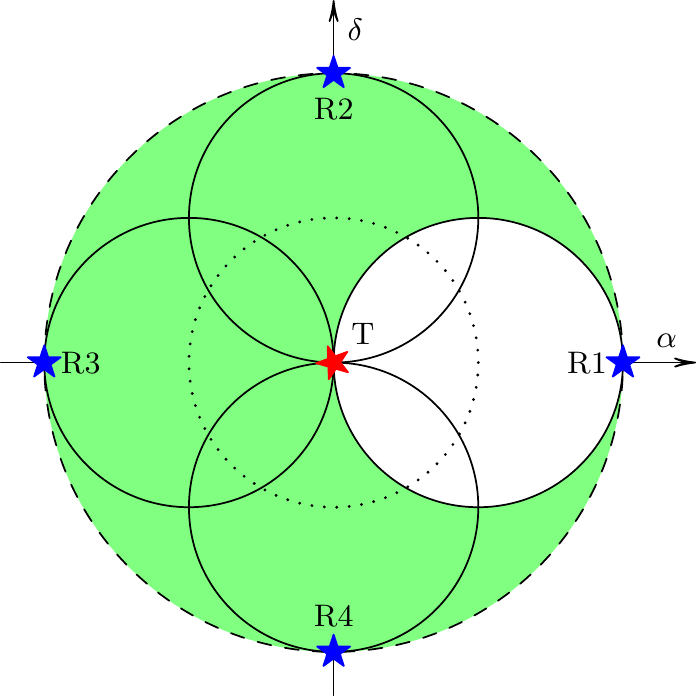} 
\caption{Field ($2^\circ$ radius, shaded) accessible to the target star $T$ 
(center) by pointing the telescope along the dotted circle ($1^\circ$ radius): 
reference stars $R1$, $R2$, $R3$ and $R4$ are shown at $\pm 2^\circ$ on either 
coordinate. 
\label{fig:principle}} 
\end{minipage}\qquad
	\begin{minipage}[c]{0.47\textwidth}
\vspace*{3mm}
	\includegraphics[width=\hsize]{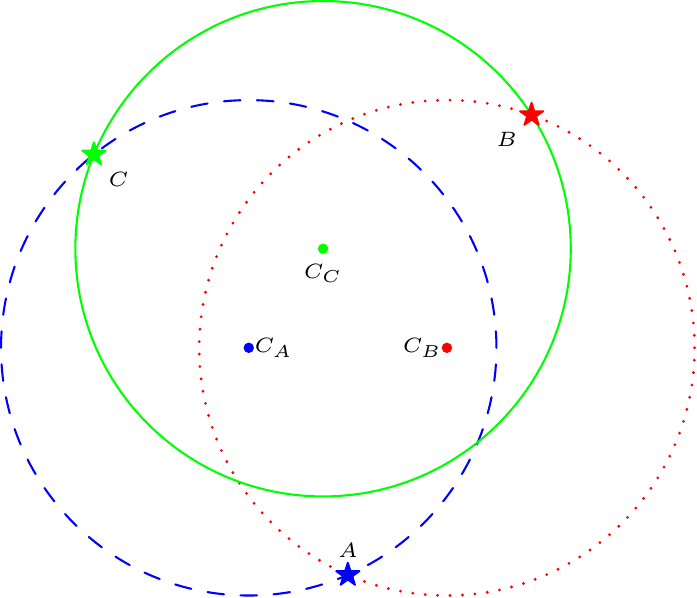} 
\vspace*{3mm}
	\caption{Example of pairwise observation of three targets $A$, $B$ and $C$, using three pointing offset positions $C_A$, $C_B$ and $C_C$ of the 
		optical axis. 
		\label{fig:3target}} 
\end{minipage}
\end{figure}

The annular field of RAFTER (radius $\theta$, width $\delta\theta$) allows 
simultaneous observation of source pairs at a significantly larger angular 
separation (up to $2 \theta$) than in case of a conventional round geometry 
detector, which concentrates 
the same area $2\pi \theta \, \delta\theta$ within 
radius $\sim \sqrt{2 \theta \, \delta\theta}$. 
The principle is depicted in Fig.\,\ref{fig:principle}, with 
$\theta = 1^\circ$ and $\delta\theta << \theta$: 
the target star $T$ is located at the centre, 
and four reference stars ($R1$, $R2$, $R3$ 
and $R4$) are placed at extreme positions on either axes. 
The target and any object in the whole shaded area (dashed circle) can be 
observed simultaneously, by setting the telescope optical axis 
in suitable points on the dotted circle, i.e. pivoting the field of view 
around the target. 
For example, a set of three targets, labelled $A$, $B$ and $C$, as shown in 
Fig.\,\ref{fig:3target}, can be observed pairwise by application of small 
pointing offsets among subsequent positions $C_A$, $C_B$ and $C_C$ of the 
optical axis. 

Such instrument concept allows selection of the actual field of view close to a 
given target with some liberty, according to performance optimisation criteria. 
In particular, we may select a region around the science target including the 
brightest accessible reference star; other {\it ad hoc} criteria may be adopted 
when convenient, e.g. maximising the number of field stars. 
Some such options are investigated in our study and detailed in 
Sec.\,\ref{Sec:simul}. 

This framework appears to be efficient toward implementation of robust 
narrow angle astrometry, exploiting available references (the Gaia 
catalogue) and/or strengthening it with additional observations to improve 
on individual source parameters. 

In Sec.\,\ref{Sec:ScienceCase} we briefly review some of the compelling science 
topics supporting a high precision relative astrometry mission; in 
Sec.\,\ref{Sec:rafter} the RAFTER instrument concept is recalled and some 
relevant aspects of operation with an annular field are outlined. 
In Sec.\,\ref{Sec:exoplanets} the practical case of relative astrometry on 
a specific exoplanetary system is expounded in some detail. 
In Sec.\,\ref{Sec:RingField} we derive the main statistical implications of 
annular field observation around the brightest stars in the sky. 
Finally, in Sec.\,\ref{Sec:conclusions}, we draw our conclusions.

\section{Overview of the science case}
\label{Sec:ScienceCase}
As shown in Sect.~\ref{Sec:rafter}, RAFTER is a suggested technical implementation of a relative astrometry mission similar e.g. to Theia. The science cases of these two missions, thus, share a lot of similarities, differences emerging especially in consideration of the different observing strategies allowed by the annular field of view of RAFTER vs. the concentrated one of Theia.

Limiting ourselves to a short review of the possible scientific applications of this mission concept, RAFTER has the possibility to contribute to the detection and orbit characterization of Earth- to Superearth-sized exoplanets, as well as to several General Relativity and Cosmology related issues. For example, high precision astrometry on selected stellar samples will be able to probe the distribution (shape, radial profile, lumpiness) of the Dark Matter (DM) halo of the Milky Way, and possibly M31, by determination of the dynamics of dwarf spheroidal galaxies (dSph), globular clusters (GC) and halo streams within the Local Group as mapped by a significant fraction of their brightest stars, and precise masses, distances and proper motions to binary stars with black hole (BH) and neutron star (NS) companions will also be obtained from $\mu as$ level measurement of their orbits. This includes the prototype of this class, Cygnus\,X-1 \citep{MillerJones21}.

Another potential contribution is in the field of Solar System dynamics. In the Solar System there exist $\sim100$ objects whose maximum visual magnitude is $V\leq10$ or brighter, for which RAFTER can provide Micro- and sub-micro-arcsecond astrometry. Most of them are asteroids of the main belt, whose ephemerides can be improved by these measurements, at least for the vast majority of them that was not visited by a dedicated space probe, with potential implications in the field of Solar System dynamics. Moreover, high-precision determination of the orbits of the Galilean satellites can be combined with data from the Juno or the planned ESA JUICE mission to obtain improved constraints on the internal structure of Jupiter.

More details on a few selected aspects of the scientific case are given in the following subsections.

\subsection{ Exoplanets: State of the Art }
\label{Sec:ExoContext}
The current exoplanet count is at level 5000. Specifically, as of 2021 November 1st, 
NASA Exoplanet Archive ({\tt exoplanetarchive.ipac.caltech.edu}) lists 4566 confirmed planets (along with 4663 TESS 
transit candidates), while the less restrictive Extrasolar Planets Encyclopaedia
({\tt exoplanet.eu}) lists 4868 confirmed in 3597 systems. 
Most such planets have been detected through photometric transit and radial velocity 
measurements ($\sim 71$\% and 20\%, respectively), while the others have been discovered 
using astrometry, transits timing, microlensing and other methods. 
\citet{Perryman18book} offers comprehensive discussion of various techniques 
employed to detect exoplanets.

The systems with confirmed planets evidence a rich diversity of objects. For example, 
planetary masses range over six orders of magnitude, extending from a 70\,$M_\mathrm{J}$ 
brown dwarf \citep{10.1093/mnras/staa1608} down to a 0.02\,$M_\oplus$ super Mercury \citep{2012ApJ...752....1R}.
A remark concerning terminology is due: although brown dwarfs are classified as 
substellar objects, we do not differentiate between them and planets for brevity. 
This should raise no ambiguity in the text. 

The current exoplanetary census contains 43 confirmed Earth-like planets, i.e. with masses 
ranging from 0.6 to 2.3\,$M_\oplus$. 
Most of them (36) have been detected from primary transit searches, while microlensing 
observations and radial velocity surveys revealed five and one planet, respectively. 
The discovery of the last planet in the list, KOI-55\,c, is remarkable: its existence 
was inferred from the brightness pulsation of its host star \citep{2011Natur.480..496C}.

The large number of detected exoplanets encourage the belief that many, if not all, stars 
host planetary systems. The major detection techniques, i.e. photometric transit and 
radial velocity, however, suffer from a strong selection effect. 
Transits are observable only in case of edge-on systems, when our line-of-sight is almost 
parallel to the orbital plane. 
The probability that orientation of a planet orbit is favorable for transit detection 
is determined 
by the ratio of the stellar radius $R_\star$ to the size of planet orbit $a_\mathrm{p}$ \citep{1984Icar...58..121B,Perryman18book}:
\begin{equation}
    p\simeq0.005\left(\frac{R_\star}{R_\oplus}\right)\left(\frac{a_\mathrm{p}}{1\,\mathrm{AU}}\right)^{-1}\,.
\end{equation}
The geometric probability depends on neither star distance nor planet size. 
For example, for a Sun twin ($R_\star = R_\odot$), the probability is 
0.5\% for the Earth and 0.1\% for Jupiter. 
For a solar type star, the probability is above 1\% if $a_\mathrm{p}\lesssim 0.5\,\mathrm{AU}$, while for a A0V host star with $2.5R_\odot$ a 1\%-probability zone extends to 0.8\,AU.

Although conditions are more relaxed for radial velocity measurements, discovery efficiency 
is still maximal for edge-on orbit orientation. 
The variation of star velocity due to orbital motion becomes less detectable as orbit 
inclination and/or companion mass decrease ($m \sin i$ effect). 

{\em Thus, only a small fraction of exoplanets can be addressed by the two most widely used 
detection techniques. } 
Face-on orbits, when orbital plane is nearly perpendicular to the line of sight, are 
mostly undetectable by either transit or radial velocity methods. 
In contrast, astrometric measurements are not subject to the above mentioned geometric limitations and can detect reflex motion of the star in any orbit configuration. 

We address astrometric exoplanet detection in more detail in Sect.~\ref{Sec:exoplanets}. 
Like radial velocities, astrometric measurements are more sensitive to massive components. 
There is a marked difference in the way that observed effects depend on star distance. 
Both eclipse depth and radial velocity amplitude are distance-independent, though the 
related photometric accuracy, of course, is affected. 
As a geometric effect, however, the astrometric signature scales inversely with distance 
(see Eq.~(\ref{eq:astrsig}) below). 
This naturally favors usage of astrometry to search for planets orbiting stars in the 
solar vicinity. 

It is worth mentioning that each detection technique has its own advantages and disadvantages; 
simultaneous exploitation of different techniques offers the best chances for determination 
of various parameters of components and host stars. 
The multi-planet system $\pi$ Mens\ae, for which combination of spectroscopic observations 
carried out at the ESO’s Very Large Telescope, photometric transits observed by TESS, and 
Gaia astrometric data provided the 3D architecture of its planetary orbits 
\citep{2020A&A...642A..31D}, gives a good examples of such synergy. 

Moreover, the combination of new and previous astrometric measurements can provide new insight 
and better parameter determination on several cases, e.g. resolving binaries with 
stellar or substellar companions by proper motion anomaly \citep{Kervella19}.

\subsection{General relativity and Cosmology-related experiments }
\label{Sec:GRexp}
%
%
One interesting possibility is that of searching for possible deviations from Newtonian dynamics in near Wide Binaries (WB). This opportunity was explored in a recent paper by \citet{2019MNRAS.487.1653B} that simulated the feasibility of detecting MOND-like astrometric signals in the orbit of Proxima Centauri around $\alpha$ Centauri A and B. This work concluded that a successful detection requires about $T=10$~years of observations with $f=3$~observations per year at an accuracy of $\sigma\leq0.33~\mu\mathrm{as}$, which is within the reach of the proposed Theia mission \citep{2017arXiv170701348T}, and thus of the RAFTER mission as it will be explained below. This duration can be reduced by increasing the cadence of the observations, as a result of a $\mathrm{SNR}\propto T^{5/2}f^{1/2}$. Moreover, the annular field of view gives the RAFTER concept a better opportunity with respect to Theia for this specific science case. As mentioned in the cited paper, in fact, basically this experiment aims at measuring the relative acceleration between Proxima and $\alpha$ Centauri. In the case of Theia, whose field of view of $0.5^\circ$ is smaller than the angular separation of $2.18^\circ$ between these two stars, this goal can be reached only by building an accurate reference frame with a sufficient number of stable reference stars, whereas the $\sim1^\circ$ radius of the annular field of RAFTER allows, in principle, to measure directly such a relative acceleration.

The actual number of existing stellar-mass black holes is an unsolved question that 
can potentially affect our understanding of galactic evolution in the early 
universe \citep{2011ApJ...738..163W}. 
Moreover, the hypothesis that Dark Matter may be constituted by Primordial Black Holes 
(PBH) in the solar-mass range had been thoroughly investigated in the past and rejected 
by several observational constraints (see e.g.\ \citet{2020JCAP...09..022J} for a list 
of relevant references). 
In the same paper, however, the author casts doubts on these conclusions, arguing that 
these limits rely on the correctness of the PBH binary semi-major and eccentricity 
distribution, and that it can evolve to be made consistent with present data. 
One technique that can be used to detect these PBH is the astrometric microlensing, 
and RAFTER can open up new possibilities in this respect.

Microlensing happens when a massive object passes in between an observer and a 
background source of light, aligned exactly or almost exactly with them. 
In such a case, the light of the background source is deflected by the massive 
objects, which behaves like a lens whose optical effects, unlike normal lenses, 
are produced by its gravity pull. 
The background object can appear magnified and distorted at the observer's position.

This phenomenon stems from the well-known effect of light deflection, stating that 
a mass $M$ shifts the observed position of a light source from its nominal position 
on an amount given by $\alpha=4GM/(c^2b)$, where $b$ is the impact parameter of the null 
geodesic connecting the source and the observer.

In the microlensing phenomenon, the original source is seen very close to the 
deflecting object by the observer, and as a consequence it is split into two 
distinct images, each presenting two main effects \citep{1998ApJ...502..538B}: 
a \emph{photometric\/} effect in which the intensities of the two images are 
magnified, and an \emph{astrometric\/} effect in which the background source 
is displaced from its nominal position by a small angle.

Quantitatively, the appearance of microlensing is governed by the lens mass and 
by the three distances between the source (s), the lens (l) and the observer (o). 
The impact parameter can be written as function of these quantities as 
$b=d_\mathrm{so}/(d_\mathrm{sl}d_\mathrm{lo})$, and in the literature the 
angular displacement is often defined in units of the so-called \emph{angular 
Einstein radius}
\begin{equation}\label{eq:einstein_radius}
    \theta_E=\sqrt{\frac{4GM}{c^2}\frac{d_\mathrm{sl}}{d_\mathrm{so}d_\mathrm{lo}}}=\sqrt{\frac{4GM}{c^2}\left(d_\mathrm{lo}^{-1}-d_\mathrm{so}^{-1}\right)},
    \label{eq:thetae}
\end{equation}
which represents the displacement when the three objects are perfectly aligned.

By defining the angular separation $u$ between the lens and the source in units of $\theta_E$, the intensity of each image writes
\begin{equation}
    A_{1,2}=\frac{u^2+2}{2u\sqrt{u^2+4}}\pm\frac{1}{2};
\end{equation}
since typically $\theta_E\sim1~\mathrm{mas}$, for on-ground observations the image separation is below the resolution of the observing instrument, and the lensed source appears as a single object with intensity
\begin{equation}
    A=A_1+A_2=\frac{u^2+2}{u\sqrt{u^2+4}}.
\end{equation}

Similarly, for what regards the \emph{astrometric\/} effect, the background source is displaced from its nominal position by an angle
\begin{equation}\label{eq:astrometric_effect}
    \delta\theta=\frac{u}{u^2+2}\theta_E.
\end{equation}

While microlensing events are relatively easy to detect with photometric measurements, its astrometric displacement is very difficult to measure, at least from the ground. On the other side, estimating the mass of the lens exploiting photometric measurements only is extremely difficult when the lens is a single objects, as in the case of a PBH.

Without entering into much detail, which the interested reader can find in \citet{2016ApJ...830...41L}, it is sufficient to note that Eq.~(\ref{eq:thetae}) shows how the mass of the lens can be inferred by the knowledge of $\theta_E$. This quantity is directly related to the astrometric displacement, while, for single-object lenses, its estimation with photometric measurements only requires a nearly direct passage of the source over the lens.

The potential of microlensing for determining trigonometric parallaxes is also worth mentioning. As recently demonstrated by \citet{2022aa.657.a18}, an intensive photometric follow-up of the microlensing event Gaia19bld allowed not only to compute the lens mass but also to derive its distance with a 10\% accuracy. Operating in continuous observing mode, RAFTER can provide dense coverage of brightness variation during a microlensing event to find lens properties from photometric data, in addition to the astrometric effect described by Eqs.~(\ref{eq:einstein_radius}) to (\ref{eq:astrometric_effect}). Thus, RAFTER is potentially capable of using both photometric and astrometric distance determination methods, giving another good example of synergy between different observational techniques.

RAFTER can help to shed light on the problem of the existence and composition of Dark Matter also by helping to improve the sample of Hypervelocity stars (HVS). These objects are considered a promising powerful probe to infer information about the gravitational potential of the Milky Way in the galactic halo, namely the Dark-Matter-dominated region of our galaxy according to the concordance $\Lambda$CDM model.

A recently published paper \citep{2019MNRAS.487.4025C} proposes a new technique able to put stringent constraints on the gravitational potential of the Milky Way using the mass, position, and velocity distributions of the HVSs. In this paper the authors put their method to test under an ideal but realistic scenario based on the predicted results of the ESA \emph{Gaia\/} astrometric mission, showing that 200 HVSs are able to provide estimations of the Navarro-Frenk-White (NFW) potential parameters with sub-percent uncertainties.

The accuracy of these constraints depends on the size of the HVSs sample, and previous works of the same authors \citep{2017MNRAS.470.1388M,2018MNRAS.476.4697M} showed that suitable HVS candidates can be identified with neural network techniques that depend solely on the astrometric parameters. The reliability of the identification, however, depends on the uncertainties on parallaxes and proper motions, a requirement that conveniently matches the RAFTER purpose of prolonging and improving the astrometric accuracy of the \emph{Gaia\/} catalog.

In addition to the improvement on the HVSs sample, RAFTER can contribute to the Dark Matter investigation also by observations of halo Wide Binaries. Their small binding energy, in fact, makes them very susceptible to external perturbations (passing stars, molecular clouds, spiral arms, large-scale tides, and massive objects in general), which allows them to be used to study the medium in which they are immersed. Therefore, halo wide-binaries can be used to study fine details of the gravitational potential of that Galactic component, placing relevant constraints on the nature of Dark matter \citep{2004ApJ...601..311Y,2010ASPC..435..453Q}. Given their extremely long periods (and large semi-major axis) their confirmation as a true gravitational pair requires very high-precision relative astrometry, currently available for only relatively small samples (see e.g., \citet{2004ApJS..150..455G,2018MNRAS.481.2970C,2020ApJS..246....4T}). Gaia DR3 will surely provide many new candidates (usually through common proper-motion, supplemented with high-precision ground-based radial velocities), but an instrument like RAFTER could be used to follow-up on them, and confirm/discard their true binary nature.

\section{ The RAFTER concept }
\label{Sec:rafter}
The telescope, recently described in the literature \citep{RAFTER_SPIE_20}, 
is designed with the 
goal of achieving good imaging quality, and above all good astrometric 
performance over a comparably large field of view. 
The latter requirement is implemented through the 
prescription of preserving circular symmetry at each stage of the optical 
system, down to and including the focal plane. 
The annular field thus provide invariance of the image characteristics along 
the azimuthal coordinate (at fixed angular radius from the optical axis), 
and smooth variation in the radial direction. 
This ensures that systematic errors related to instrument optical response 
are minimised, at least along one coordinate. 
The Roman Space Telescope (RST) also uses part of the annular corrected field 
of its optical design \citep{Pasquale14} in its Wide Field Imager, whose 
$3 \times 6$ chips are arranged in a semi-lunar (``smile") layout. 

The proposed design achieves sub-$\mu as$ systematic error on azimuth, even 
with moderate perturbations to the nominal configuration, and $< 2\,mas$ 
distortion over the radial range $[0^\circ.9, 1^\circ.1]$, easily modelled. 
The diffraction limited field ($>1.25$\ square deg) is partially populated 
by 66 $4k$ detectors on a $4'$ wide ring ($0.26$\ square deg). 

\subsection{ Optical system main characteristics }
\label{Sec:instrument}
\begin{figure}[ht!]
	\centering 
\includegraphics[width=0.9\hsize]{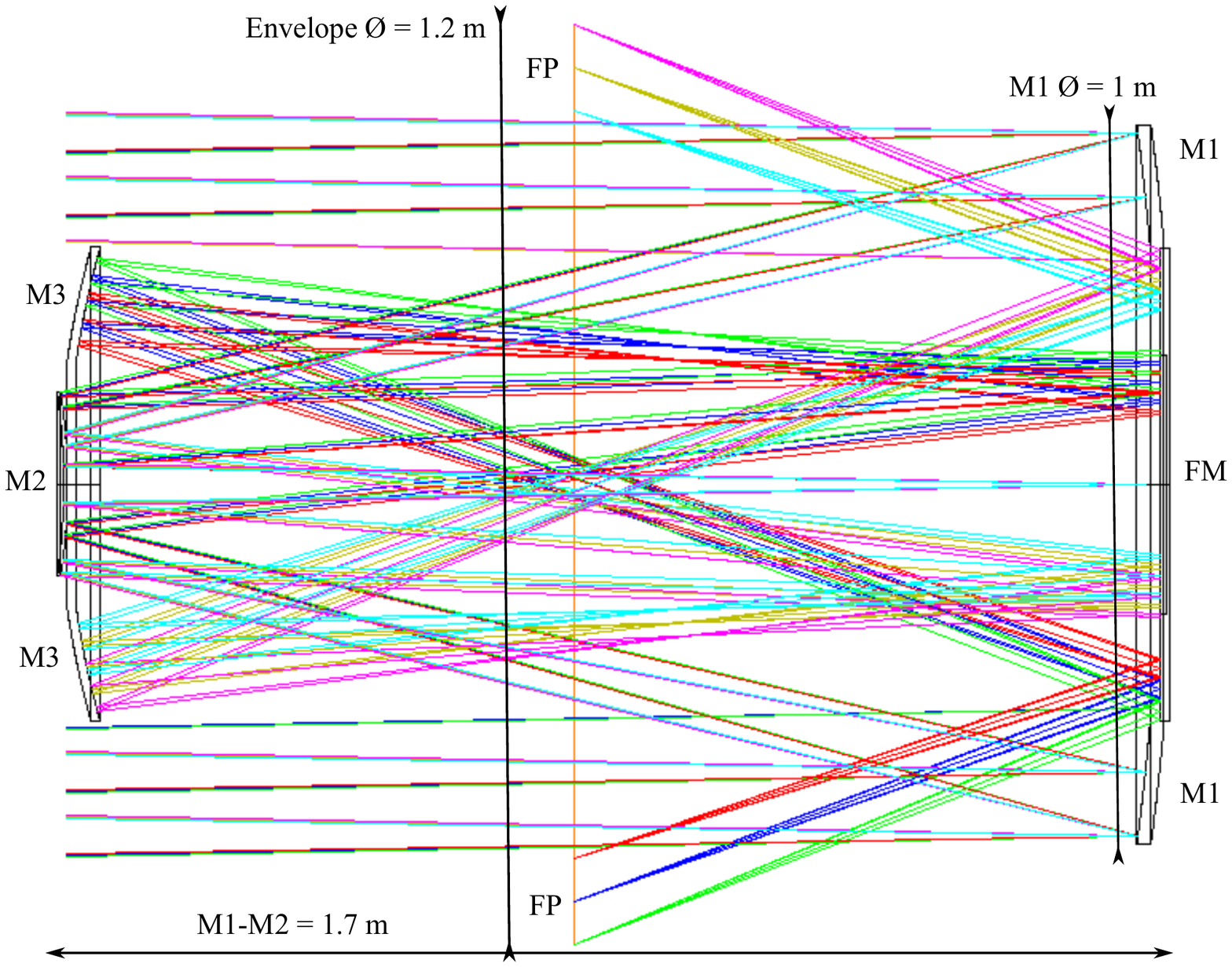} 
\caption{The RAFTER telescope layout. The focal plane (FP) is deployed on a ring surrounding the input beam (baffling not shown).\label{fig:telescope}} 
\end{figure}

The RAFTER design is derived from a classical Three Mirror Anastigmat (TMA) 
\citep{Korsch77}, optimised with the explicit goal of exploiting a 
full ring within the annular corrected field. 
The telescope layout is shown in Fig.\,\ref{fig:telescope}, in its 
CCD-compatible version with effective focal length $EFL = 30\, m$. 
The peculiarity is that the secondary (M2) and tertiary (M3) mirrors are placed 
within each other, as well as the primary (M1) and a flat folding mirror (FM). 
The input pupil and all mirrors are annular, i.e. they use only a circular 
region. 
The flat FM has the double function of feeding the optical path from 
M2 to M3, and from M3 to the focal plane (FP), a ring of CCD detectors around 
the telescope main tube. 
The system is therefore highly symmetric and very compact: $1.7\ m$ length, 
$1.2\ m$ diameter. 
Optical design characteristics, including indications on robustness against 
perturbations, have been published \citep{RAFTER_SPIE_20}. 

\subsection{ A strawman mission: key elements }
\label{Sec:mission}
The actual design of payload and satellite is beyond the scope of the current 
study, mostly focused on the instrument concept and related operation options. 
A few aspects are proposed, in order to define the framework for the subsequent 
discussion. 

RAFTER is assumed to be operated on a pointed satellite, aimed at comparably 
long observations of selected targets to achieve high precision, rather than a 
scanning mission covering the whole celestial sphere (or large fractions thereof), 
and necessarily devoting little time to individual sources. 

The $1\,m$ class telescope is consistent with the main payload of a medium class 
mission, driving satellite and operation design. 
Additional payloads are not considered herein, but such option will be an 
obvious concern of the supporting agencies throughout early mission definition 
phases. 
The operating orbit can be selected with some freedom after finalisation of the 
science case priorities; GEO and L2 are obvious possibilities suggested by previous 
missions. 
The mission lifetime is assumed to be of order of five years. 
Longer duration, obviously impacting costs, is beneficial to an astrometric 
mission, allowing observation of more targets and/or improved determination 
of their dynamical parameters, above all for long period orbits. 

At any given time, the sky area accessible will be limited by attitude 
consideration, including the need to avoid the Sun on the payload, and to 
feed the solar array providing the power supply. 
Besides, this results in a geometric constraint on observations, so that 
most targets may be observed a few times a year, e.g. in epochs grouped over 
about a semester. 

We assume a typical observation time of order of one hour. 
If the fractional time required to switch between subsequent targets is 
$\sim 10\%$, it may be possible to operate order of 20 observations per day, 
for a total of $\sim 7,000$ observations per year. 
Since several sources will require two or more observations per year, 
an acceptable number of individual science targets is on the order of few 
thousand. 

Careful definition of the science case will therefore be required to maximise 
the mission output, trading off the competing requirements of large samples 
and fine sampling over the mission lifetime. 
In practice, a few targets, requiring higher precision and/or time resolution, 
may be observed several times, and others less frequently.

\subsection{ High cadence observations }
\label{Sec:HiCadence}
Broadband observation of bright sources with a sizeable telescope results in 
fast detector saturation. 
A long integration, required to achieve high precision thanks to a correspondingly 
high photon budget, must therefore be split into many shorter elementary exposures, 
with the data co-added either on board or on ground. 
This also provides a valuable high frequency information on pointing 
stability, which may be taken advantage of by the on-board 
attitude and orbit control system (AOCS). 
Multiple exposure astrometry on bright stars is also discussed in a separate 
paper (Gai et al., submitted to PASP). 

Full detector readout on a short timescale would be impractical, but, since bright 
sources are few and usually wide apart, we can set windows around them for 
selective readout, discarding less interesting regions. 
On-board detection, required for proper management of readout windows, is routinely 
used on Gaia. 

The detector readout is assumed to be performed on a pixel period of $5\,\mu s$, 
not very challenging with respect to modern science instrumentation. 
Full frame readout on single-output $4k \times 4k$ devices requires about 80\,s, 
thus window readout is required for all chips imaging sources brighter than 
$G \simeq 15.5\, mag$. 
Several large windows (e.g.\ $44\times44$ pixels) can be read with 
source magnitude $G \gtrsim 8\, mag$; then, size and number of windows 
must be reduced. 
A single, small window ($10\times10$ pixels) per chip may allow imaging for 
stars as bright as $G \simeq 2.5\, mag$. 



\subsection{ Calibration hints }
\label{Sec:calibration}
The capability of an annular field with radius $\theta$ to observe at the same time 
a selected target and any source within angular distance $2\theta$ was anticipated 
in the Introduction and illustrated in Fig.\,\ref{fig:principle}. 

By rotation around the optical axis, the same pair of objects can be observed 
repeatedly in different detector positions, thus contributing to both 
measurement statistics and calibration. 
Subsequent estimates of the pair separation can be expected to differ by amounts 
related to the instrument response variation over the field, assuming that 
the observation is short with respect to the natural time scale of variation 
of both payload and astrophysical sources. 
Repeated observations on longer time scales will remove the degeneration between 
correlated instrument variations, and uncorrelated evolution of different 
sources. 

The goal precision of RAFTER is below the error level expected on individual sources 
in the final Gaia catalogue, but comparable to the intrinsic precision of the 
underlying Gaia reference frame. 
Therefore, statistics on a sufficiently large sample will average down the 
resulting collective error to adequate levels. 

Pairwise observation of several nearby objects, not simultaneously fitting 
within the annular field of view, allows measurement of the target against 
different reference sources and with different sampling geometry with respect 
to the instrument radial and azimuthal coordinates, as in Fig.\,\ref{fig:3target}. 

\subsection{ Relative astrometry: the local reference frame }
\label{Sec:RelAstrom}
Each target is located with respect to the available population of field stars, 
materialising a local representation of the global Gaia reference frame. 
Since the goal measurement precision is below the $1\,\mu as$ range, we may 
wonder whether the Gaia precision is adequate to the task. 

Although a complete answer for each science case should be substantiated 
by a thorough, dedicated investigation, 
we address briefly some of the main aspects involved. 
Simple conceptual considerations and simulations seem to provide encouraging 
hints with respect to the feasibility of our ambitious precision goal. 

The catalogue uncertainty on reference objects affects the evaluation of any 
target's dynamics in different ways: 
\\ 
{\bf - individual position errors} induce an offset on the photocenter, 
constant over the mission lifetime. 
\\ 
{\bf - individual proper motion errors} induce a continuous photocenter drift, 
appearing as a contribution to the linear motion of the target. 
\\ 
{\bf - individual parallax errors} generate a reflex parallax term on the 
target. 

The first term is inessential for most purposes, and the second term does not 
affect the measurement of the target's astrometric wobbling. 
The third contribution is potentially more critical, since it induces an 
apparent oscillation on the target's linear motion, which may be erroneously 
interpreted as the astrometric effect of an orbiting companion. 

Of course, astrometric measurements are intrinsically affected by troubles 
in the detection of orbital periods close to one year, because it is difficult 
to disentangle them from the modulation applied by the Earth's motion 
around the Sun. 
The difficulty is somewhat mitigated by geometry, since the target's orbit will 
not usually be similar to the Earth's orbit projection involved in the 
parallax method; however, the superposition might make the estimate more prone 
to errors. 
The degeneration may be reduced by a long sequence of astrometric observations, 
and sometimes broken by introduction of additional astrophysical information. 

\subsection{ Cosmic noise }
\label{Sec:cosmic}
A concern for high precision astrometry is the intrinsic astrometric variability 
of reference stars around a selected target, induced by a number of reasons: 
photospheric activity, starspots, undetected companions, and so on. 
While the issue is quite relevant and deserves {\em ad hoc} studies for any 
specific application and target, we can remark that the above assessment 
on the number of field stars provides some encouraging element. 

Each source may have its own amplitude and time scale of variation; however, 
they can hardly be expected to have the same direction and phase. 
The mitigation of catalogue errors by averaging them over a sufficiently large 
sample of field stars (as in Sec.\,\ref{Sec:CatError}) is a sort of ``brute force" 
statistical approach. This may be expected to work as well in levelling out the 
individual astrometric fluctuations from the photo-center estimate used as a reference 
to evaluate the target motion.

\section{ Astrometry optimization on Exoplanets }
\label{Sec:exoplanets}
Astrometric information is used in exoplanetary studies to solve two tasks: 
discovery of new objects and characterization of known systems. Both these tasks are based 
on examining effect of exoplanets, or, more generally, unseen companions, on their host 
stars. Detection of planets relies on analysis of deviation of apparent path of the host 
from the single stars. Size of this effect is conveniently described by the so-called 
astrometric signature \citep{2018MNRAS.476.5658B, Perryman18book, 2018A&A...614A..30R, Sozzetti+2014}
\begin{equation}\label{eq:astrsig}
 \upsilon=3\ \mu\mathrm{as}\times 
  \left(\frac{M_\mathrm{p}}{M_\oplus}\right)
  \left(\frac{M_\star}{M_\odot}\right)^{-1}
  \left(\frac{a_\mathrm{p}}{1\ \mathrm{AU}}\right)
  \left(\frac{d}{1\ \mathrm{pc}}\right)^{-1}\,,
\end{equation}
where $d$ is the distance, $a_\mathrm{p}=\left(M_\star/M_\mathrm{p}\right)a_\star$ is the semi-major 
axis of the planet orbit,  $M_\star$ and $M_\mathrm{p}$ the host star and 
planet mass, respectively; $a_\star$ is the size of the stellar orbit around the system's barycenter.

For Earth-type planets, astrometric signatures are at the sub-$\mu$as 
level. 
For example, if a $1\,M_\oplus$ planet orbits a Solar-mass star at 
$a_\mathrm{p}=1\ \mathrm{AU}$, the signature ranges from 0.3 to 0.03 $\mu$as 
for stars in the volume between $d=10$ and 100 pc. 
In contrast, massive planets have more prominent effect on their host stars. 
For instance, the impact of a Jupiter ($a_\mathrm{p}=5\ \mathrm{AU}$)  amounts 
to $\simeq 1$\ mas for a $0.5\,M_\odot$ M dwarf at 10\ AU. 
However, planets at large $a_\mathrm{p}$ need relatively long measurements 
for reliable detection. 

\subsection{ Mission related constraints }
\label{Sec:MissionConstr}
Simulations showed that astrometry is most efficient in discovering exoplanets 
with orbital period not exceeding the duration of observations
\citep{Perryman+2014,Sozzetti+2014}. 
In view of the anticipated mission lifetime of 5\ yr, therefore, we restrict 
our analysis to planets with orbital period $P\lesssim 5$\ yr. 
Making use the Kepler's third law, we obtain the corresponding constraint 
on the detectable orbit size 
\begin{equation}\label{eq:maxsize}
    a_\mathrm{p}\lesssim a_\mathrm{max}\simeq 3\ \mathrm{AU}\times\left(\frac{M_\star}{M_\odot}\right)^{1/2}\,.
\end{equation}
This condition means that for upper main sequence host stars 
($M_\star\simeq 10\,M\odot$) the semi-major axis of planet orbit is 
limited to $\lesssim 10$\ AU, while it should be below $\simeq 1$\ AU 
for lower main sequence stars ($M_\star\simeq 0.1\,M\odot$).

Astrometric discovery of exoplanets crucially depends on the signal-to-noise 
ratio $\upsilon/\sigma$, with $\sigma$ being the astrometric accuracy. 
We adopt the ordinary three-sigma rule, that is, we assume 
$\upsilon\ge 3\sigma$ in the following. 
This criterion, together with the above condition for the maximum orbit 
size, enables us to establish a lower limit to the planet mass. 
Rearranging Eq.~(\ref{eq:astrsig}) and applying Eq.~(\ref{eq:maxsize}), 
we find that
\begin{equation}\label{eq:minmass}
 M_\mathrm{p}\gtrsim M_\mathrm{min}\simeq 
  M_\oplus
  \left(\frac{\sigma}{3\ \mu\mathrm{as}}\right)
  \left(\frac{M_\star}{M_\odot}\right)^{1/2}
  \left(\frac{d}{1\ \mathrm{pc}}\right)\,.
\end{equation}
This equation is convenient for estimation of the minimum mass of 
astrometrically detectable planet in terms of accuracy, host star 
mass and distance.

Our simulations (Sec.~\ref{Sec:FieldPrec}) show that a precision level 
of few $\sigma = 0.1\ \mu$as is achievable with the RAFTER design on 
one hour exposures on very bright stars ($G \lesssim 4 \ mag$); 
the value $\sigma = 0.2\ \mu$as is used below, in our assessment of 
planets that can be discovered. 
Table~\ref{tab:StarStat} gives predictions for the accuracy, including 
the precision level expected on nearby reference stars materialising the 
reference frame. 
However, these estimations are affected by operations and 
other error sources, 
which are preliminarily addressed in Sec.~\ref{Sec:FieldPrec} to 
\ref{Sec:CatImpro}. 
Therefore, the value of 0.2\ $\mu$as is to be taken as a representative 
estimate of foreseen astrometric performance. 

\begin{figure}[ht!]
	\centering 
	\begin{minipage}[c]{0.47\textwidth}
\includegraphics[width=\hsize]{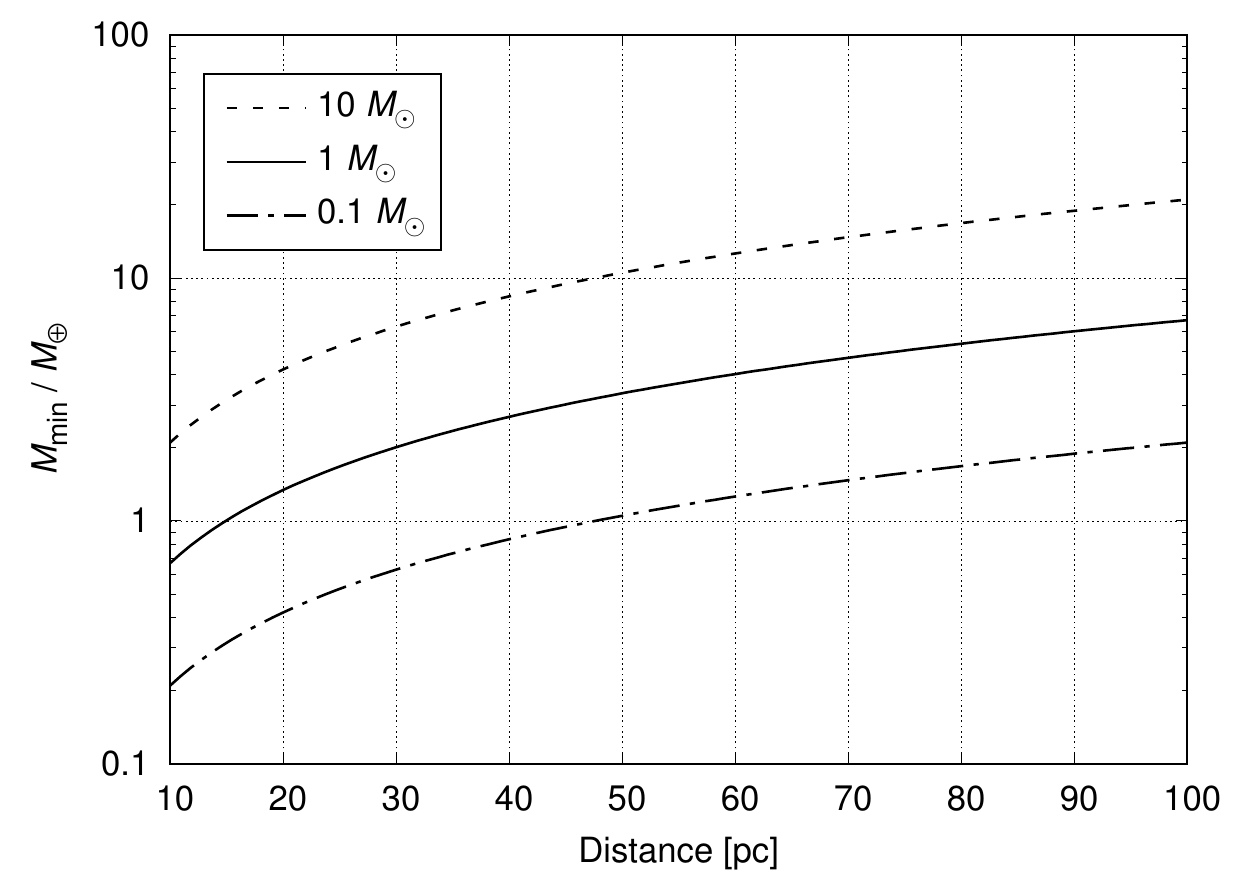}
\caption{ Minimum mass of detectable planet as a function of distance. 
The estimation assumes astrometric $SNR = 3$. The solid, dashed and 
dash-dotted lines correspond to a solar-mass, upper- and lower-main 
sequence host star, respectively.\label{fig:minmass}}
\end{minipage}\qquad
	\begin{minipage}[c]{0.47\textwidth}
	\includegraphics[width=\hsize]{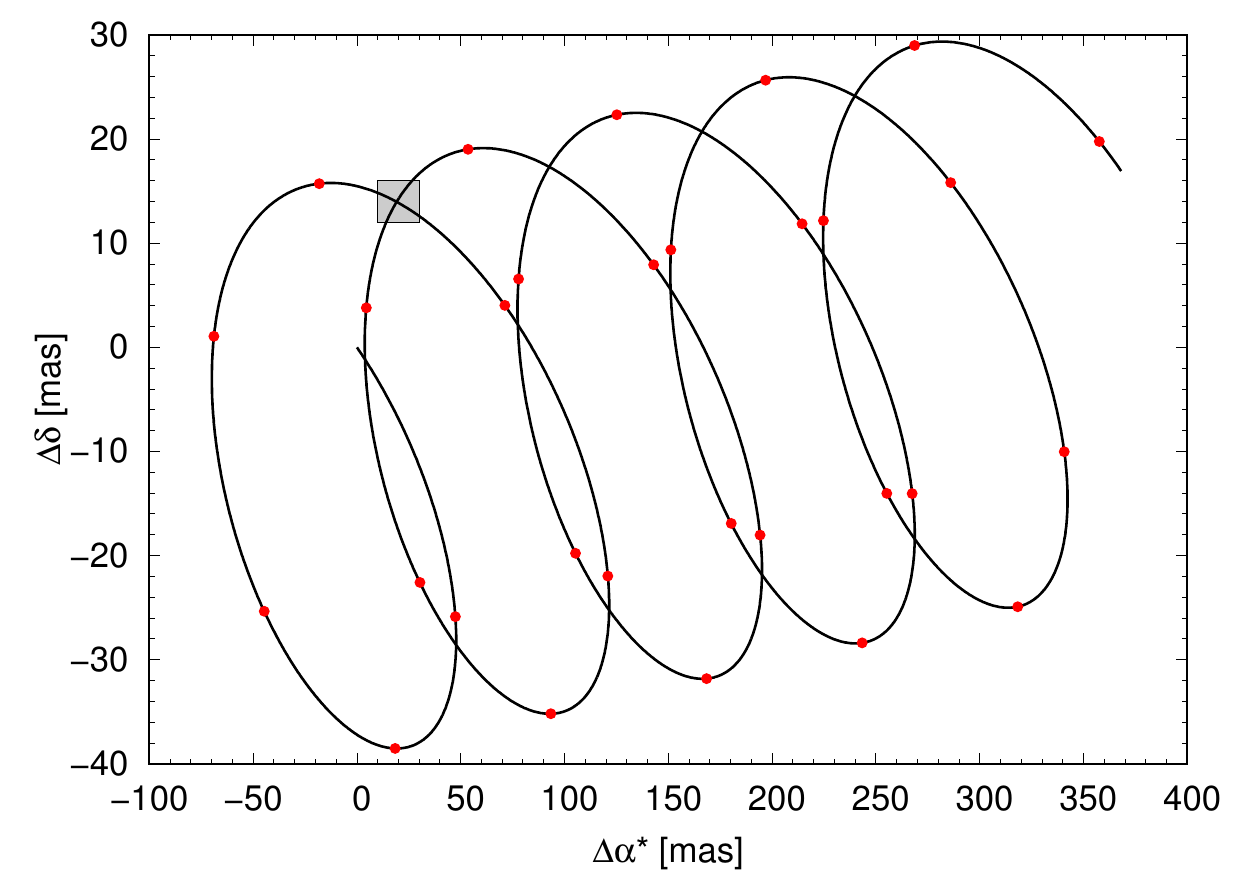}
	\caption{ Path on the sky of HD147513, over 5 yr. The coordinates are reckoned from 
		the starting position at J2026.0. The red dots give an example of observations sampled 
		at the rate of 6 observation sets per a year. The shaded rectangle indicates the 
		vicinity of a crossing point shown in Fig.~\ref{fig:HD147513b_zoom}. The area size 
		is greatly exaggerated in this drawing.\label{fig:HD147513b}}
\end{minipage}
\end{figure}

Fig.~\ref{fig:minmass} illustrates the minimum detectable planet mass 
versus distance for three different host star masses: solar ($1\, M_\odot$), 
low- ($0.1\,M_\odot$) and upper- ($10\,M_\odot$) main sequence. 
This plot shows that an Earth-mass planet can be detected up to $d=15$\ pc 
for solar-mass stars, while for $0.1\,M_\odot$ stars , e.g.\ M6 dwarfs, 
the Earth discovery limit is at $\simeq 47$\ pc. 
For a $1\,M_\odot$ host star, the lower detectable planet mass runs from 
0.7 to $7\,M_\oplus$ as distance goes from 10 to 100\ pc. 
For low-mass stars, the corresponding lower limit ranges from 0.2 
to $2\,M_\oplus$.

For massive stars, the astrometric effect from an Earth orbiting a 
$10\,M_\odot$ star remains undetectable by RAFTER even at 10 pc. 
This result lends itself to a straightforward interpretation. 
For such a star, the minimum detectable orbit size from 
Eq.~(\ref{eq:maxsize}) is $a_\mathrm{p}\lesssim 9.5$\ AU. 
The corresponding astrometric signature of an Earth-mass planet at 10 pc 
is $\simeq 0.3\ \mu$as, i.e.\ a detection level of $1.5\sigma$, 
well below the adopted $3 \sigma$-threshold. 
The minimum detectable planet mass runs from 2 to $20\,M_\oplus$, over distances 
ranging from 10 to 100\ pc, for a $M=10\,M_\odot$ host star. 
Eq.~(\ref{eq:minmass}) suggests that, at 10 pc, the astrometric motion caused 
by an Earth-mass planet is detectable for $M_\star\lesssim 2.2\,M_\odot$, 
i.e.\ for A2V and later main-sequence star. 

It is worth stressing that the detection limits refer to a mission lifetime 
of 5\ yr, scaling as a function of such parameter. 

\subsection{ Pinpointing exo-planets }
\label{Sec:Refine}
We now briefly discuss how high-precision relative astrometry can be used 
for characterization of known exoplanetary systems. 
Many exoplanets have been discovered from radial velocity measurements. 
This technique can determine only the combination $a_\star \sin i$, 
where $i$ is the orbit inclination. 
Accordingly, the planet mass is uncertain by the unknown factor $\sin i$. 
\\ 
In contrast, astrometry is capable of determining the inclination, together 
with the other Keplerian elements, and, therefore, it can provide valuable 
complementary information, allowing full characterization of planetary 
systems.

As an example, we consider HD147513b, a Jupiter mass planet orbiting a G3/G5V 
star with $M=0.92\ M_\odot$ \citep{2004A&A...415..391M}. 
This system belongs to the Gaia primary sources, and its astrometric data 
have high accuracy, in particular the parallax relative error is 
$\sigma_\varpi/\varpi = 8.5\times10^{-4}$. 
The system characteristics are summarised in Table~\ref{tab:HD147513}, 
deriving the astrometric parameters and radial velocity from the Gaia 
EDR3 \citep{Gaia-CollaborationEDR32020} and DR2 \citep{2018yCat.1345....0G}, 
respectively, while the orbital elements and the host star data are from 
the Extrasolar Planets Encyclopaedia.\footnote{http://exoplanet.eu/}

With parallax $\varpi = 77.565\ mas$, the distance to the system is 12.9 pc. 
Because of the inclination uncertainty, only the lower limit to the planet 
mass is known: $M_p\gtrsim1.21\,M_\mathrm{J}=385\,M_\oplus$. 
This, in turn, results in an astrometric displacement due to the orbital 
motion at the level of at least 130\ $\mu$as. 

In terms of our reference accuracy (0.2\ $\mu$as), this value is associated 
to an astrometric signal-to-noise ratio $\mathrm{S}/\mathrm{N}\simeq650$ 
(with maximum value of the inclination, $i=90^\circ$), 
or better with smaller inclination. 
A full astrometric solution, requiring several samples per orbit, can 
constrain the inclination, planetary mass and other parameters. 

\begin{table*}
    \centering
    \caption{Parameters of HD147513b (Gaia source ID 6018047019138644480)\label{tab:HD147513}}
    \begin{tabular}{cccc}
    Parameter & Value  & Uncertainty & Unit \\
    \hline 
    $\alpha$ & 246.0058026652  & 0.057 & deg; mas \\
    $\delta$ & -39.1929655698  & 0.038 & deg; mas \\
    $\varpi$ & 77.565  &  0.066 & mas \\ 
    $\mu_{\alpha*}$ & 73.748 & 0.082 & mas yr$^{-1}$ \\
    $\mu_{\delta}$  & 3.367 & 0.059 & mas yr$^{-1}$ \\
    $v_r$ & 12.889 & 0.132 & km s$^{-1}$ \\
    Apparent magnitude, $V$ & 5.37 & & mag \\
    Mass, $M_\mathrm{p}\sin i$ & 1.21  && $M_\mathrm{J}$ \\
    Semi-major axis, $a_\mathrm{p}$ & 1.32  && AU  \\
    Orbital period, $P$ & 528.4  &  6.3 & day \\ 
    Eccentricity, $e$  & 0.26 & 0.05 \\
    Argument of pericenter, $\omega$ & 282  & 9 & deg \\
    Pericenter passage epoch, $T_\mathrm{p}$ & 2451672.0  & 11 & JD \\
    \end{tabular}
\end{table*}

Fig.~\ref{fig:HD147513b} illustrates the apparent path of HD147513 on the 
celestial sphere, computed for the period from 2026 to 2031, using the data 
in Table~\ref{tab:HD147513}. 
The trajectory shows a looping motion, i.e. the path intercepts itself 
several times at different times. 
It is clear from simple geometrical considerations that this happens when 
the parallax is larger than the annual proper motion, hence the feature is 
common to most nearby stars. 
The time distribution of the crossing events depends on the relationship 
between parallax and absolute value of proper motion 
$\mu=(\mu_{\alpha*}^2+\mu_\delta^2)^{-1/2}$. 
For HD147513, with $\varpi=77.565$\ mas and $\mu=73.825$\ mas\,yr$^{-1}$, 
the crossings are separated by 0.84, 1.29 and 1.62 yr. 

The crossing points are favorable for planet discovery and orbit 
characterization by relative astrometry. 
Indeed, {\em when a star returns to one such region, it fits in the same 
local frame}, just slightly deformed due to proper motion and parallax 
of the reference (field) stars. 
Thus, repeated transits offer better opportunities to detect and measure 
the peculiarities of the target's motion, compared to other fields crossed 
just once by the star. 

Therefore, the observation schedule can be optimised to take advantage of 
crossing conditions. The red dots in Fig.~\ref{fig:HD147513b} exemplify 
a distribution of observations with a uniform 6 point/yr cadence, suited 
to conventional full astrometric solution, still approximately hitting 
eight of the twelve crossing points. 

\begin{figure}[ht!]
	\centering 
    \includegraphics[width=0.6\hsize]{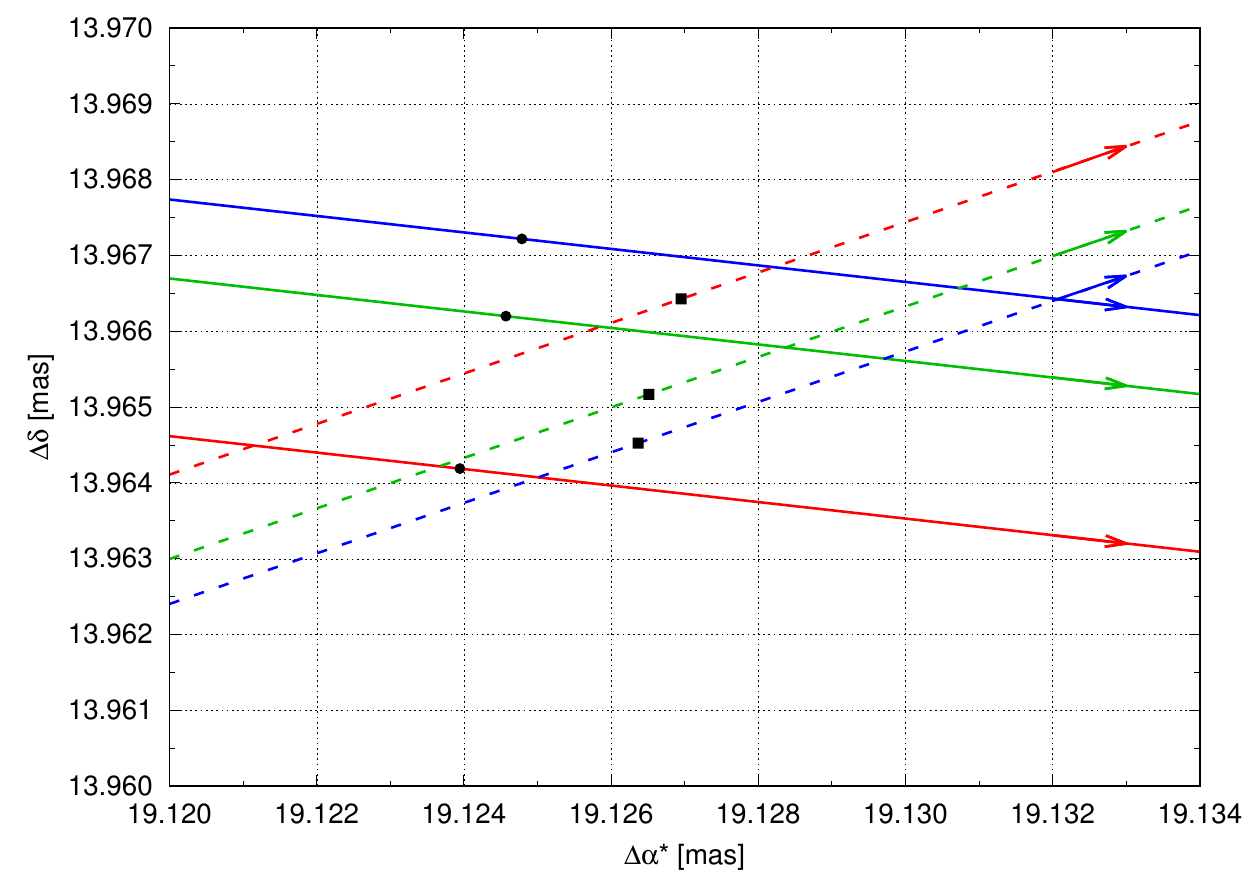}
    \caption{Zoomed view of the $14\times10\ \mu$as area marked in Fig.~\ref{fig:HD147513b}. 
    The moments of observation when the star crosses the same area are separated by 0.84 yr. 
    The red, blue and green lines show the path of the host star of the HD145513 system on 
    the sky for inclination $i=30$, 60 and 90 deg, respectively. The first passage is 
    drawn with the solid lines, while the dashed lines represent the second passage. 
    The arrows indicate the direction of motion. The black dots illustrate positions of 
    the star calculated at a fixed moment of time during the first passage. The black 
    squares exemplify the same for the second passage.\label{fig:HD147513b_zoom}}
\end{figure}

The fine structure of orbits close to a crossing point is further illustrated 
in Fig.~\ref{fig:HD147513b_zoom}, where the stellar path is shown for three 
different values of inclination; the two visits are separated by a time 
elapse of 0.84 yr. 
Different inclinations imply different planet masses: 
the mass is minimum (1.21\,$M_\mathrm{J}$) for an edge-on orbit with $i=90^\circ$, 
while larger 
values are related to smaller inclinations: 1.40\,$M_\mathrm{J}$ for 
$60^\circ$, and 2.42\,$M_\mathrm{J}$ for $30^\circ$. 
The plot shows that, for the first passage, the green line is separated 
from the blue and red lines by about 1 and $3\ \mu$as, respectively, and 
by some 50\% less in the second transit. 
This difference is due to the changing orientation of the host star orbital 
position relative to the line-of-sight. 

Using the same set of reference stars, differential astrometry is expected 
to localize the stellar path at sub-$\mu$as precision. 
This, in turn, should make it possible to estimate the inclination to 
order of $6^\circ$ ($1 \sigma$) in one crossing, because inclination affects 
the apparent stellar path. 

This example evidences the relative astrometry capability to efficiently break 
the well-known degeneracy between planet mass and inclination, with few 
measurements. 
However, a full astrometric solution can provide all seven Keplerian 
elements, provided good enough observations \citep{Perryman18book}. 
With its very high astrometric accuracy, therefore, RAFTER has a good 
potential for characterization of known planetary systems, for constraining 
planet masses as well as for improvement or independent determination 
of three-dimensional orbits. 

\begin{table}
    \centering
    \begin{minipage}[c]{0.47\textwidth}
    \caption{Reference stars for HD147513b within $0^\circ.5$, down to 
    $G = 10\,mag$
    \label{tab:HD147513_refCen}}
\vspace*{3mm}
    \begin{tabular}{cccc}
    Distance & G & RA & Dec \\
    (deg) & (mag) & (deg) & (deg) \\
    \hline 
      0.373 &   9.623 & 246.359 & -38.941 \\
      0.390 &   8.728 & 246.321 & -38.889 \\
      0.421 &   9.092 & 246.047 & -38.773 \\
      0.439 &   8.812 & 246.539 & -39.344 \\
    \end{tabular}
\end{minipage}\qquad
%
    \begin{minipage}[c]{0.47\textwidth}
    \caption{Reference stars for HD147513b within $2^\circ$, down to 
    $G = 7\,mag$. 
    \label{tab:HD147513_refAnn}}
    \vspace*{1mm}
    \begin{tabular}{cccc}
    Distance & G  & RA  & Dec \\
    \ [deg] & [mag]  & [deg]  & [deg] \\
    \hline 
      0.714 &   6.110 & 245.136 & -39.431 \\
      0.851 &   6.621 & 245.061 & -39.630 \\
      1.630 &   5.394 & 246.132 & -37.566 \\
      1.777 &   6.408 & 245.781 & -37.425 \\
      1.882 &   6.679 & 245.916 & -41.074 \\
    \end{tabular}
\end{minipage}
\end{table}

\section{ Observations with an annular field }
\label{Sec:RingField}
With respect to a conventional contiguous circular or rectangular field, an 
annular field observes at a given time several stars at some distance from a 
target (located within the focal plane ring populated by detectors), and does 
not ``see" simultaneously many other nearby sources (inside the annulus). 
Such limitation can be overcome by a set of partially overlapped 
observations, as depicted in Fig.\,\ref{fig:3target}. 
Some of the peculiar differences of annular and conventional contiguous 
field observations are evaluated below, by comparison of three cases: 
\begin{enumerate}
\item a circular field with diameter $0^{\circ}.5$, comparable that 
proposed for Theia ($\sim 0.2$~square deg); 
\item a thin annular field with diameter $2^{\circ}$ and width $2'$,
radius $58' \le \rho \le 1^\circ$, with approximately the same area; 
\item a thick annular field, width $4'$ ($56' \le \rho \le1^{\circ})$, 
corresponding to RAFTER, with area $\sim 0.4$~sq. deg. 
\end{enumerate}

With respect to the source HD147513b, described in Sec.\,\ref{Sec:Refine}, 
we have reference stars available in either case in significantly different 
magnitude ranges. 
In case\,1, within $0^{\circ}.5$ from the target, we have four stars, down 
to magnitude $G = 10\,mag$, as listed in Table\,\ref{tab:HD147513_refCen}. 
However, in cases\,2 and 3, there are five stars within $2^\circ$ distance 
brighter than $G = 7\,mag$, listed in Table\,\ref{tab:HD147513_refAnn}. 
The annular field makes available brighter reference stars, with a gain 
of about $2\,mag$ in this case. 
Such kind of evaluation is performed in our analysis to all bright targets 
($G \le 8\,mag$) in the sky, building up some relevant statistical results.

We select a sample of bright sources, and nearby reference 
stars, as described in Sec.\ \ref{Sec:simul}. 
The main aspects considered in our analysis, based on the general considerations 
anticipated in Sec.\ \ref{Sec:intro}, are: 
\begin{enumerate}
    \item 
the brightest accessible reference star (Sec.\ \ref{Sec:BriPart}); 
    \item 
the number of available bright reference stars (Sec.\ \ref{Sec:FieldStars}); 
    \item 
the photon limited uncertainty on the position of the photo-center of 
the set of reference stars, against which the target is located 
(Sec.\ \ref{Sec:FieldPrec}); 
    \item 
the uncertainty on target motion due to the limited knowledge on reference 
stars, i.e. the catalogue induced errors (Sec.\ \ref{Sec:CatError}); 
    \item 
the potential improvement on the catalogue data of the observed sample of 
reference stars achievable by exploitation of the new astrometric measurements 
(Sec.\ \ref{Sec:CatImpro}); 
    \item 
the benefits of a larger sample, achieved by setting a slightly fainter limiting 
magnitude for the reference stars (Sec.\ \ref{Sec:DeeperMag}). 
\end{enumerate}

In our assessment, all above aspects are considered independent of each other, for 
simplicity. 
In a practical case, observations should more appropriately be solved for {\em both} 
target and field stars, thus actually improving astrometry on all objects. 
Calibration is an intrinsic part of the process. 
The issue of such an ``holistic" approach will be further investigated in 
future studies.

\subsection{ Sources from Gaia EDR3 }
\label{Sec:simul}
The annular field performance is evaluated by simulation, using the
Gaia EDR3 catalogue for the astrometric parameters and magnitude 
of our star sample. 
The target stars are the 62,723 bright objects down to
$G=8\,mag$, whereas field stars are limited to $G=12\,mag$ 
(about three million objects) for most of the current exercise. 

For each target, the field center is selected according to the
criterion of including the brightest accessible partner star 
compatible with each case of focal plane geometry. 
The initial search is focused on a circular region with radius 
corresponding to the maximum detector size, as from 
Fig.\ \ref{fig:principle}. 
In case 1, such radius is $0^{\circ}.5$,
whereas in cases 2 and 3 it is $2^{\circ}$, thus providing access
to a sky area 16 times larger (the ring width is mostly irrelevant).

The field of view is then placed according to the selected target-bright
partner (BP) pair. In the central field case, the centre is placed
in the mid point of the line joining the pair, whereas in the annular
cases the optical axis is displaced sufficiently to place both stars 
in the annulus, i.e. at $1^{\circ}$ radius (see Fig.\,\ref{fig:3target}). 
Further optimization of the observing region may depend on additional science
requirements (e.g. including or avoiding specific field stars), but
the subject is not addressed in the current simple framework. 
The total number of accessible field stars is listed in 
Table~\ref{tab:StarStat}, together with some of the results of 
our analysis (described in the following). 

\begin{table*}
    \centering
    \caption{Statistics of the stellar sample; data from Gaia EDR3, $G \le 12\,mag$ 
    \label{tab:StarStat}}
    \renewcommand{\arraystretch}{1.2} 
    \begin{tabular}{lccc}
     & Central  & Annular 1 & Annular 2\tabularnewline
    \hline 
    Total number of accessible field stars  & 1,310,469 & 1,380,708 & 2,683,834 \\
    Bright partner median magnitude {[}mag{]}  & 7.390 & 5.011 & 5.011 \\
    Bright partner RMS magnitude {[}mag{]}  & 1.235 & 1.027 & 1.027 \\
    Individual bright partners          & 48,667 &  9,123 & 9,123 \\ 
    Median number of field stars, BP pointing & 16 & 17 & 32 \\
    RMS number of field stars, BP pointing & 17.2 & 16.6 & 32.4 \\
    Median number of field stars, scan & 59.0 & 233.5 & 467 \\
    RMS number of field stars, scan & 63.2 & 233.0 & 465.9 \\
    Field cumulative precision, median {[}$\mu as${]} & 0.224 & 0.155 & 0.143 \\
    Field cumulative precision, RMS {[}$\mu as${]} & 0.112 & 0.106 & 0.064 \\
    \end{tabular}
\end{table*}

\subsection{ Brightest partner matching each target }
\label{Sec:BriPart}
The field of view can be selected to include the target and the brightest 
accessible field star, hereafter Bright Partner (BP), considered as the most 
precise available astrometric reference thanks to its favourable photon budget. 

\begin{figure}[ht!]
	\centering 
    \includegraphics[width=0.48\textwidth,height=0.29\textheight]{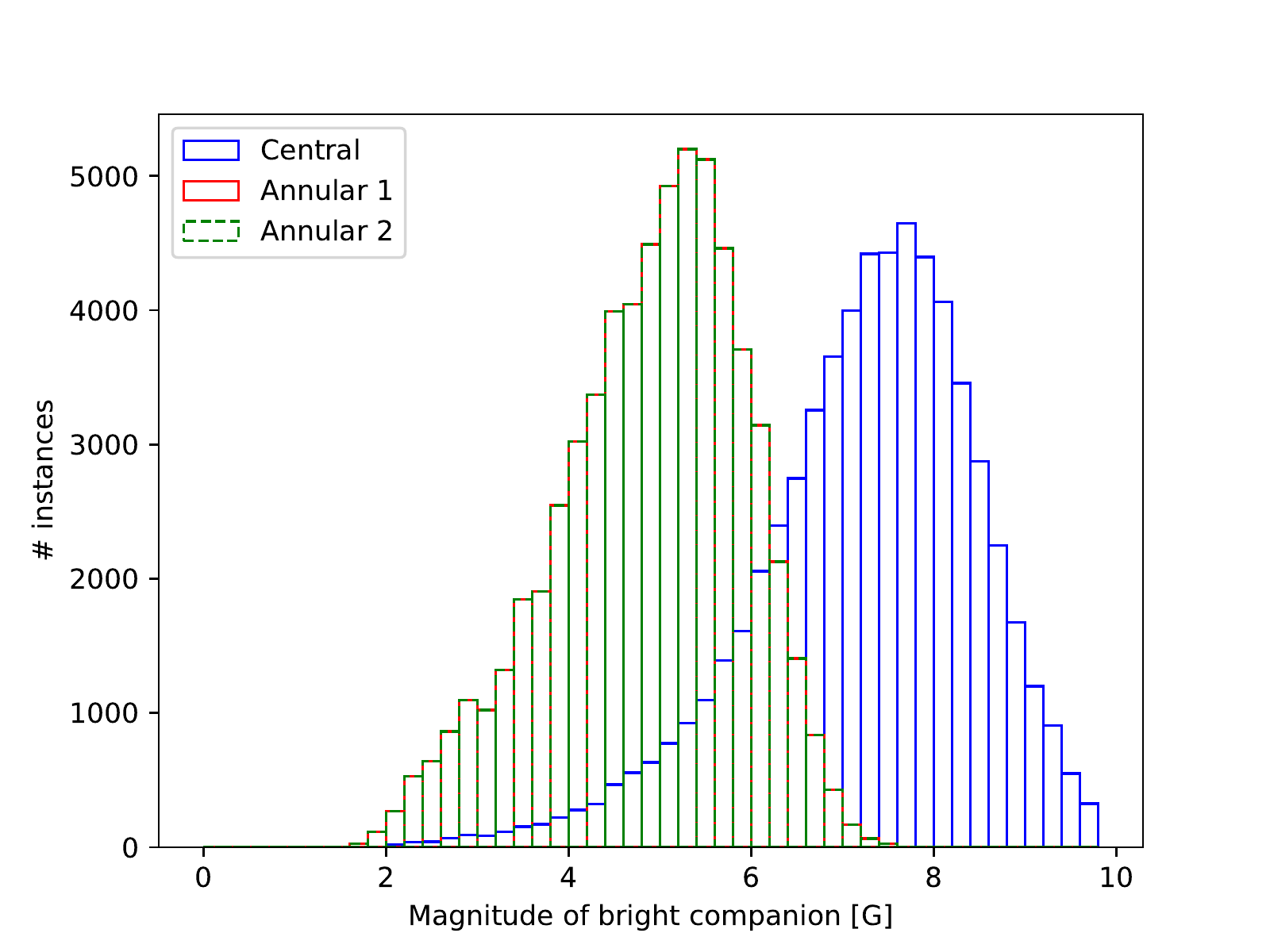}
	\includegraphics[width=0.48\textwidth,height=0.29\textheight]{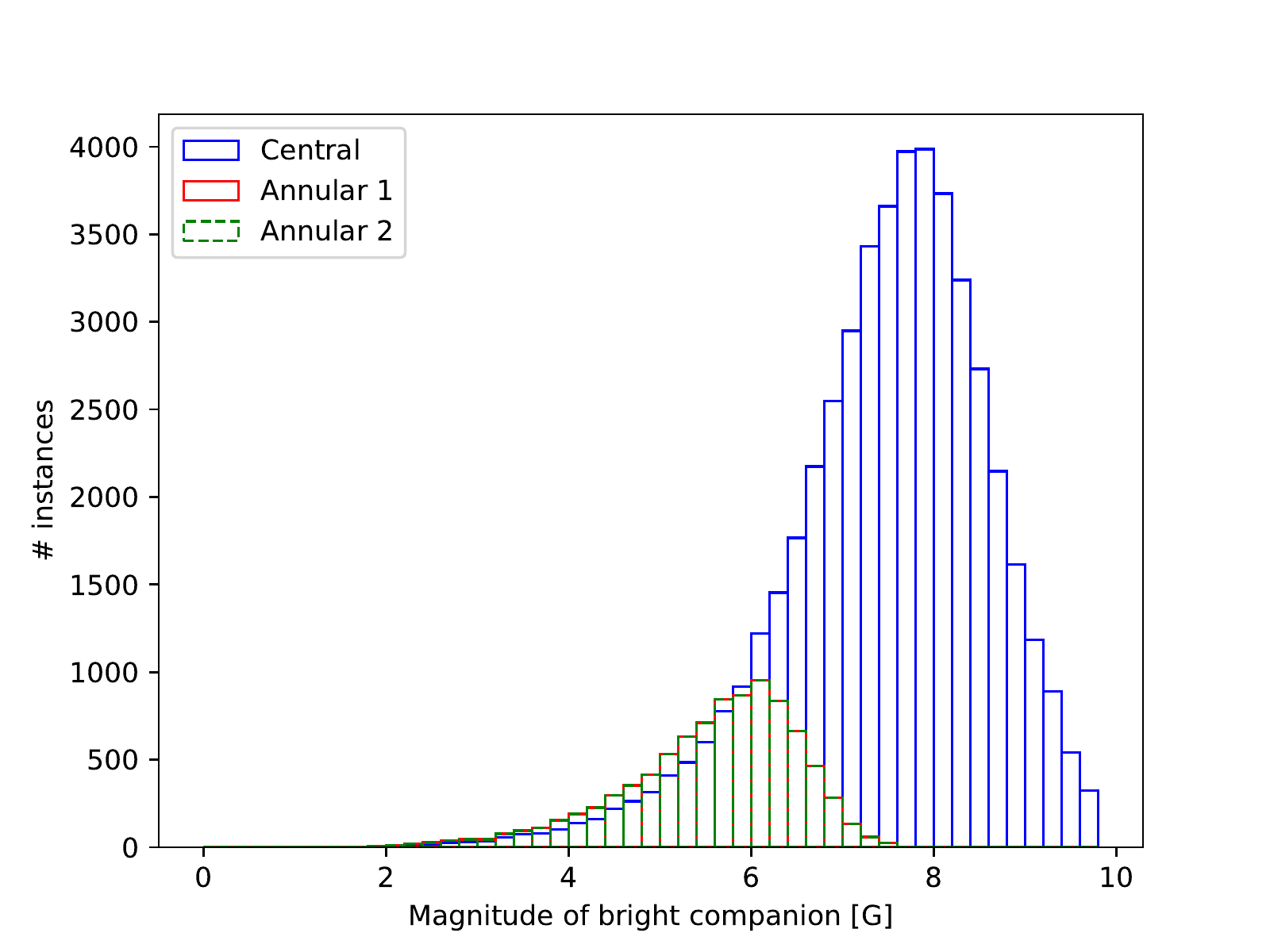}
    \caption{Bright partner magnitude: collective (one per target, left) and individual (duplicates removed, right).\label{fig:BriPart}}
\end{figure}

For each target, the BP magnitude is brighter in the annular field
cases (2 and 3) than for the central field by about two magnitudes,
as shown in Fig.~\ref{fig:BriPart}. 
The number of BPs is larger than the number 
of stars at a given magnitude (left panel in figure), because 
several nearby targets are paired to a common reference bright star 
by the field selection strategy. 
Conversely, the central field case must resort to a larger number of fainter BPs 
in the close neighbourhood of the target (right panel). 
This results in a higher sampling of bright stars over the set of
measurements, used as BP to several fainter targets. 
The median and RMS values of BP magnitude are listed in 
Table~\ref{tab:StarStat}. 

Due to the annular fields definition, the same bright targets and BPs are
selected in both cases 2 and 3. 

\subsection{ Field scan for more reference stars }
\label{Sec:FieldStars}
An alternative observing strategy is considered, in order to evidence 
the annular field flexibility, consisting in selecting different fields 
for each target (by pointing offset, see Fig.\,\ref{fig:principle}) 
throughout observations, thus piling up larger sets of 
reference stars. 
The two observing strategies will be referred to in the following as 
``BP pointing" and "scan", respectively. 

Releasing the requirement of simultaneous observation of target and BP, it is 
possible to point the instrument with more flexibility within the accessible 
area, thus increasing the number of observed field stars. 
E.g., we may assume three observations per year over five years (nominal 
mission lifetime), totalling 15 visits uniformly 
distributed in time, placed with a pointing offset such that the accessible 
region around each target is ``scanned" in a roughly uniform way. 
This is beneficial e.g. in averaging down the catalogue errors affecting the 
determination of each target's motion, as discussed 
in Sec.\ \ref{Sec:CatError}. 

\begin{figure}[ht!]
	\centering 
    \includegraphics[width=0.48\textwidth,height=0.29\textheight]{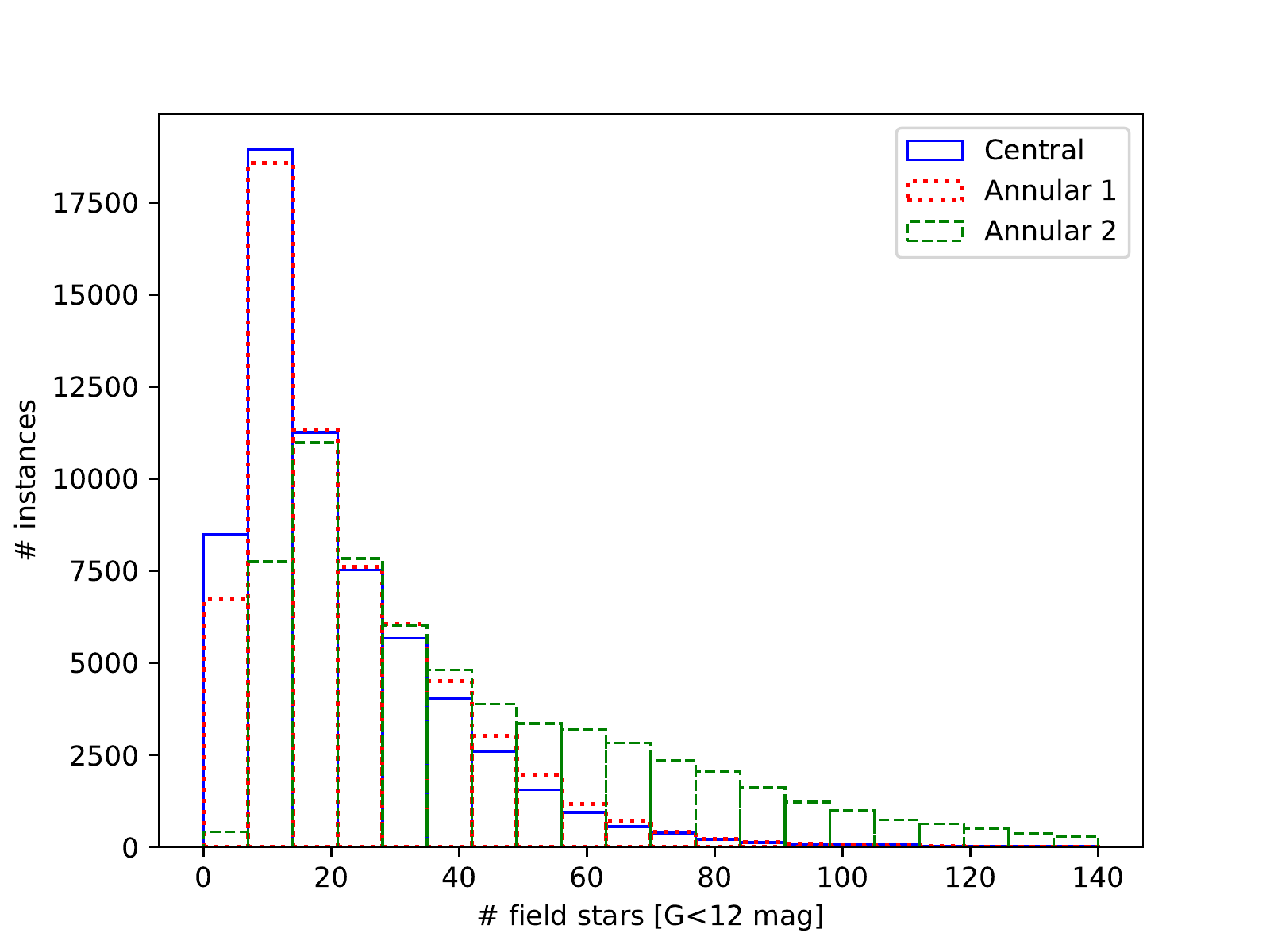}
	\includegraphics[width=0.48\textwidth,height=0.29\textheight]{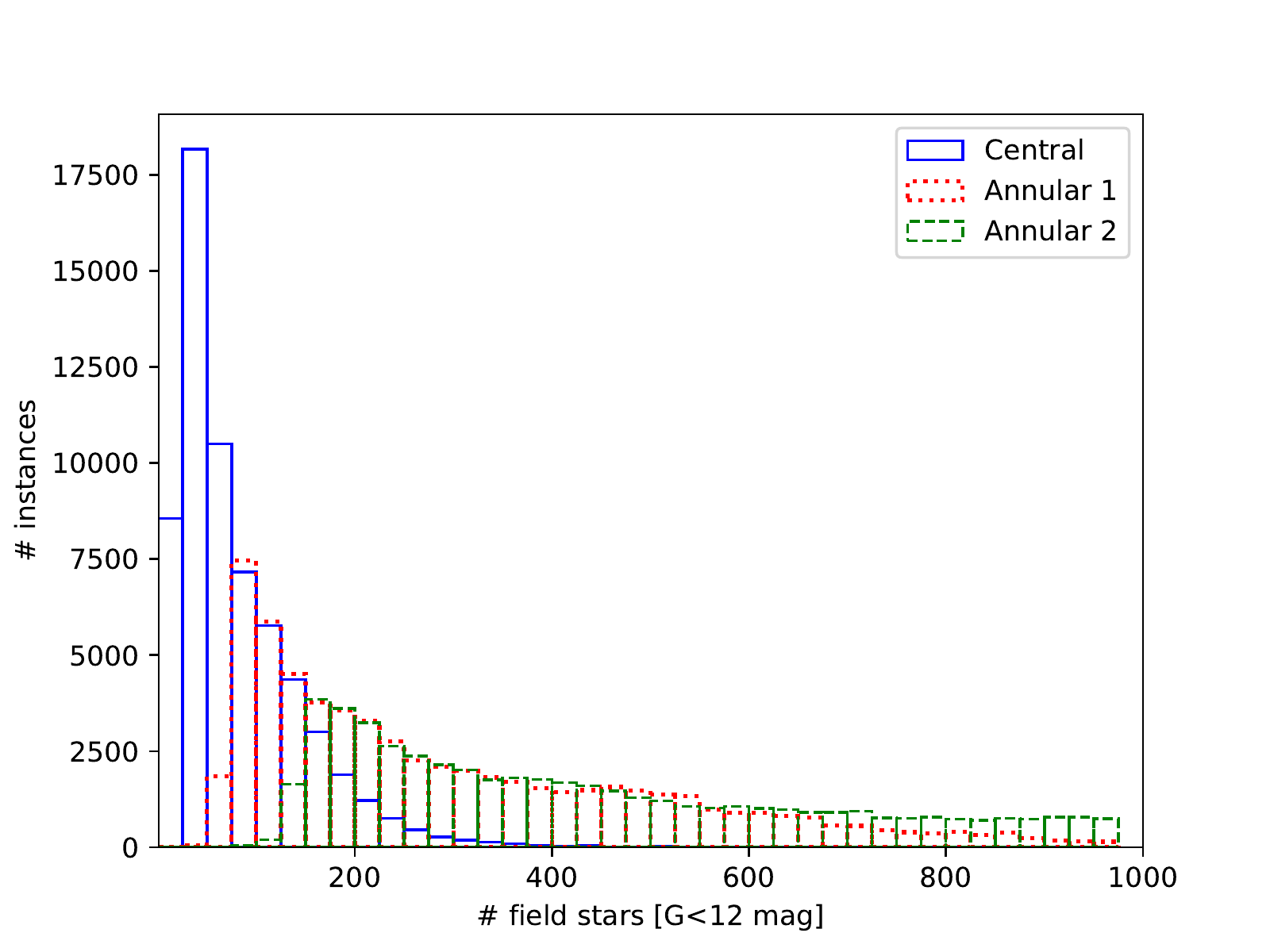}
    \caption{ Histogram of field star population within $G=12\,mag$, BP pointing 
    (left) and field scan (right) observations. 
    \label{fig:FieldStars}}
\end{figure}

The histogram of the number of field stars down to $G=12\,mag$, for BP  
pointing, is shown in Fig.~\ref{fig:FieldStars} (left). 
Both central field and thin annular field cases provide
comparable distributions (blue and red lines), according to their
equal sky area coverage, with a median value of 16 and 
17 reference stars, respectively, whereas the thick annular field case (green 
line) features a significantly larger number of field stars ($>30$), thanks 
to its larger area. 

Using multiple pointing offsets, 15 in our case, it is possible to 
accumulate observations over more field stars, within an accessible area which 
is larger for the annular than for the central field case. 
The resulting population of reference stars increases in any case, but the 
central field reaches a median star count $\sim 60$, against $>230$ and 
$>460$ respectively for the thin and thick annular field cases. 
The statistics is evidenced by the histograms in Fig.~\ref{fig:FieldStars} 
(right), and some of the resulting relevant figures are listed in 
Table~\ref{tab:StarStat}. 

\subsection{ Photon limited uncertainty on reference frame }
\label{Sec:FieldPrec}
The field stars represent a materialization of the Gaia reference
frame, providing a ``grid'' against which the target's motion 
can be evaluated. 
The impact of catalogue errors will be discussed in Sec.\,\ref{Sec:CatError}; 
here we just address the photon limited precision associated to the 
whole set of field stars selected for each target. 
For the current exercise, we assume BP pointing, with one hour exposure; 
the results from field scan are obviously expected to be even better, depending 
on implementation. 

The target position with respect to the set of $N$ field stars can
be defined by its angular separation to each of them, whichever 
their actual current positions.
The location uncertainties $\sigma_{n},\,n=1,\ldots,$N associated to each
star's magnitude is therefore combined by weighted average to provide the 
collective photo-center uncertainty $\sigma_{C}$ (target excluded): 
\begin{equation}
    \frac{1}{\sigma_{C}^{2}} = \sum_{n=1}^{N}\frac{1}{\sigma_{n}^{2}}\,.
\label{Eq:FieldCumError}
\end{equation}
This can be considered as the measurement precision of the reference frame 
in the current observation (single epoch). 

\begin{figure}[hb!]
	\centering 
    \includegraphics[width=0.48\textwidth,height=0.26\textheight]{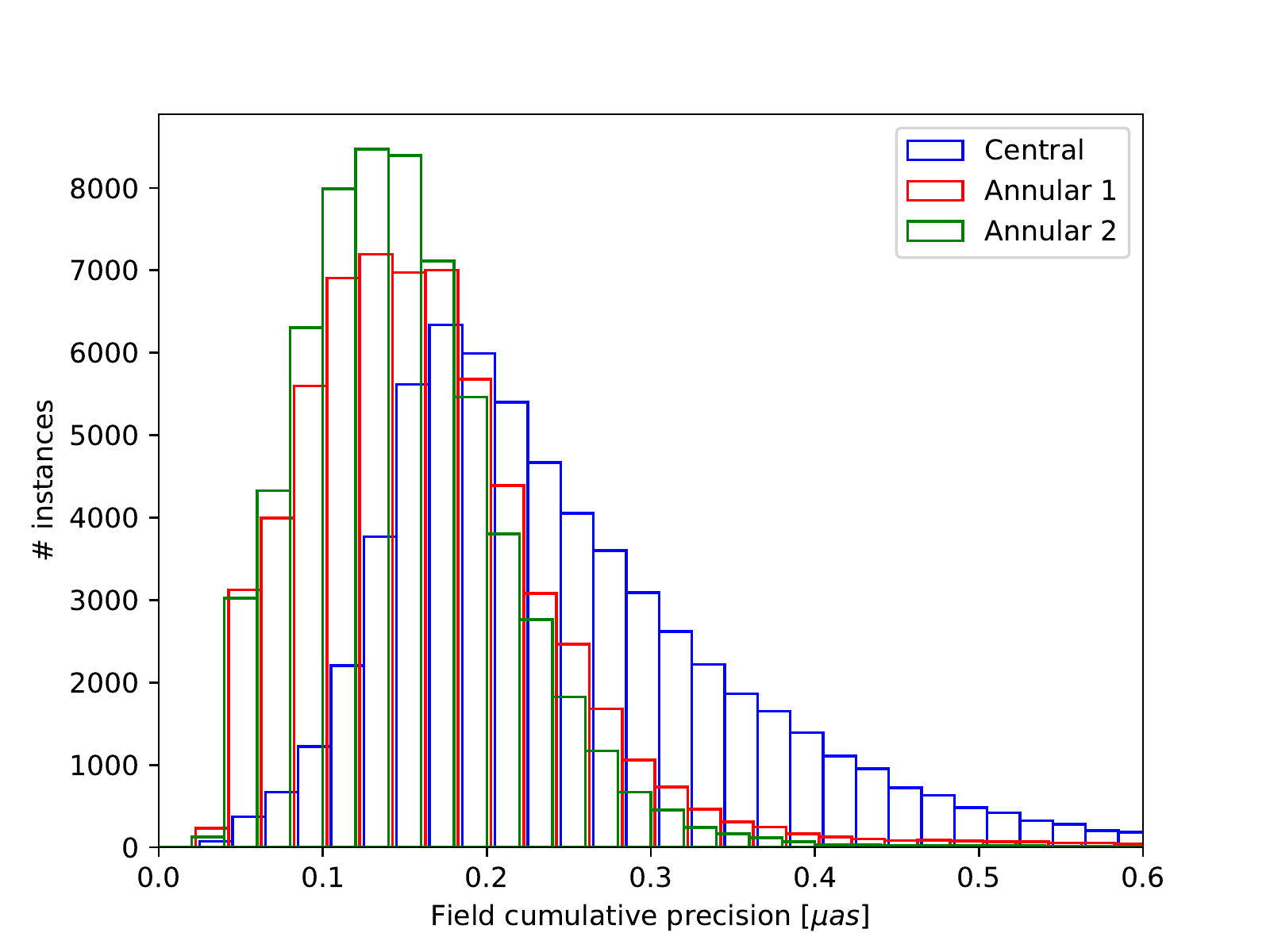}
	\includegraphics[width=0.48\textwidth,height=0.26\textheight]{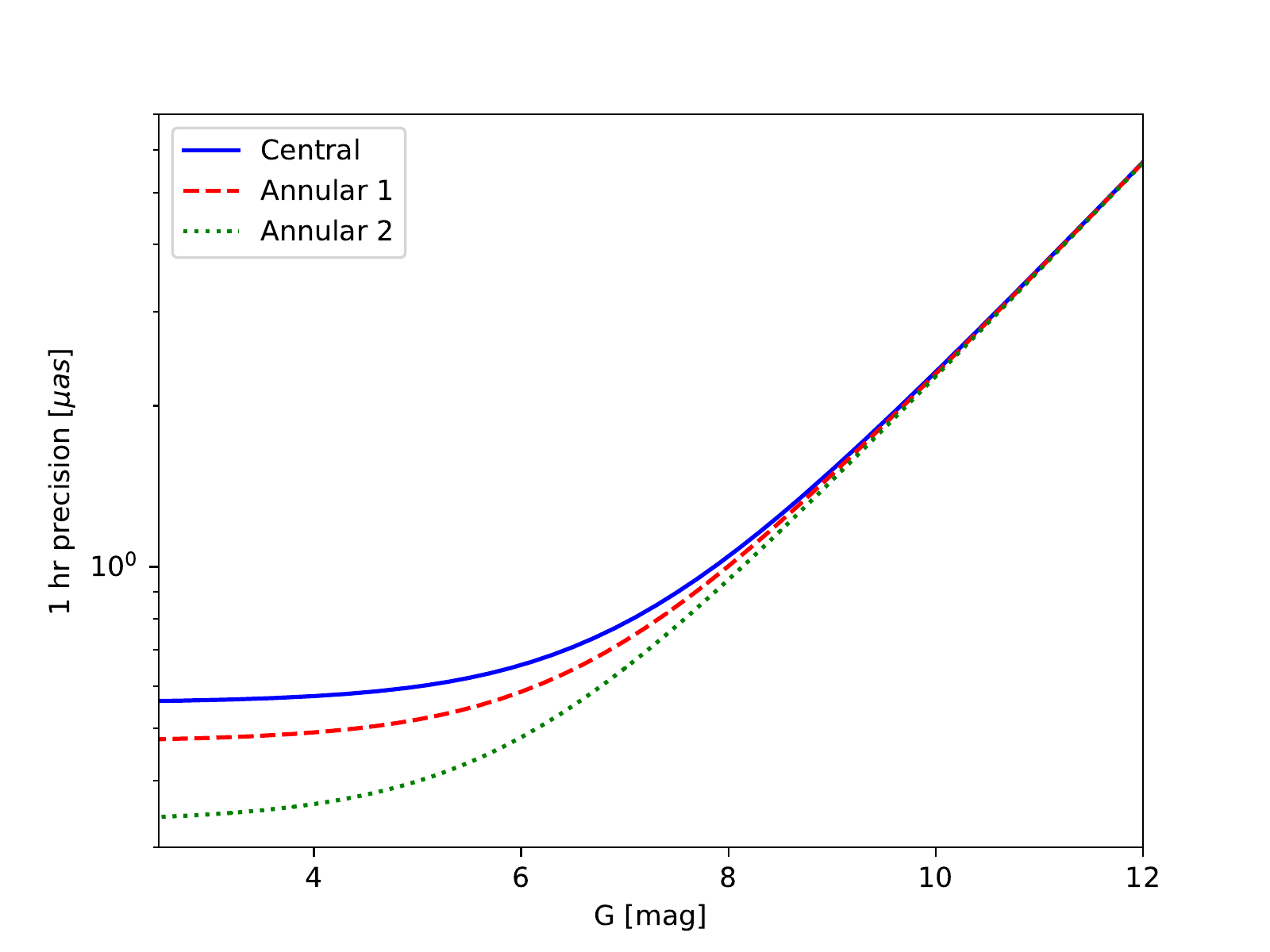}
    \caption{ Cumulative field photocenter precision (left) and individual epoch 
    	precision, including field photocenter error (right).
    	\label{fig:FieldPrec}}
\end{figure}

For each target, the above collective uncertainty $\sigma_{C}$ is computed, 
and its statistics is considered. 
The distribution of the results, shown in Fig.~\ref{fig:FieldPrec} (left), 
is for most targets below $1\,\mu as$ for the central
field case (blue line), and it improves progressively for the thin
(red line) and thick (green line) annular field cases. 
The annular fields also evidence a more compact distribution, dominated 
by the higher precision of the lower magnitude BP. 

The precision associated to individual observations is based on the photon 
limited precision \citep{Gai2017PASP,Mendez2014} for unresolved stars, 
diffraction limited images, and considering a $30\%$ degradation factor from 
realistic instrument disturbances, for an exposure time of one hour per epoch. 
This is composed with a calibration error related to the field photocenter 
precision, choosing as a conservative value the median value of the distribution 
(Fig.\,\ref{fig:FieldPrec}, left) $+3\sigma$, which includes $>98\%$ of the cases. 
The resulting individual measurement uncertainty is shown in 
Fig.\,\ref{fig:FieldPrec} (right); the solid blue line refers to the central field, 
the dashed red line to the thin annular field, and the dotted green line to the 
thick annular field. 


Assuming other error sources are kept at bay, sub-$\mu as$ precision is achieved 
for point-like sources brighter than $G \simeq 8\,mag$; the noise floor is 
therefore dominated by the field star statistics, and can be improved 
with observing strategy (e.g. field scan). 

\begin{figure}[ht!]
	\centering 
    \includegraphics[width=0.48\textwidth,height=0.26\textheight]{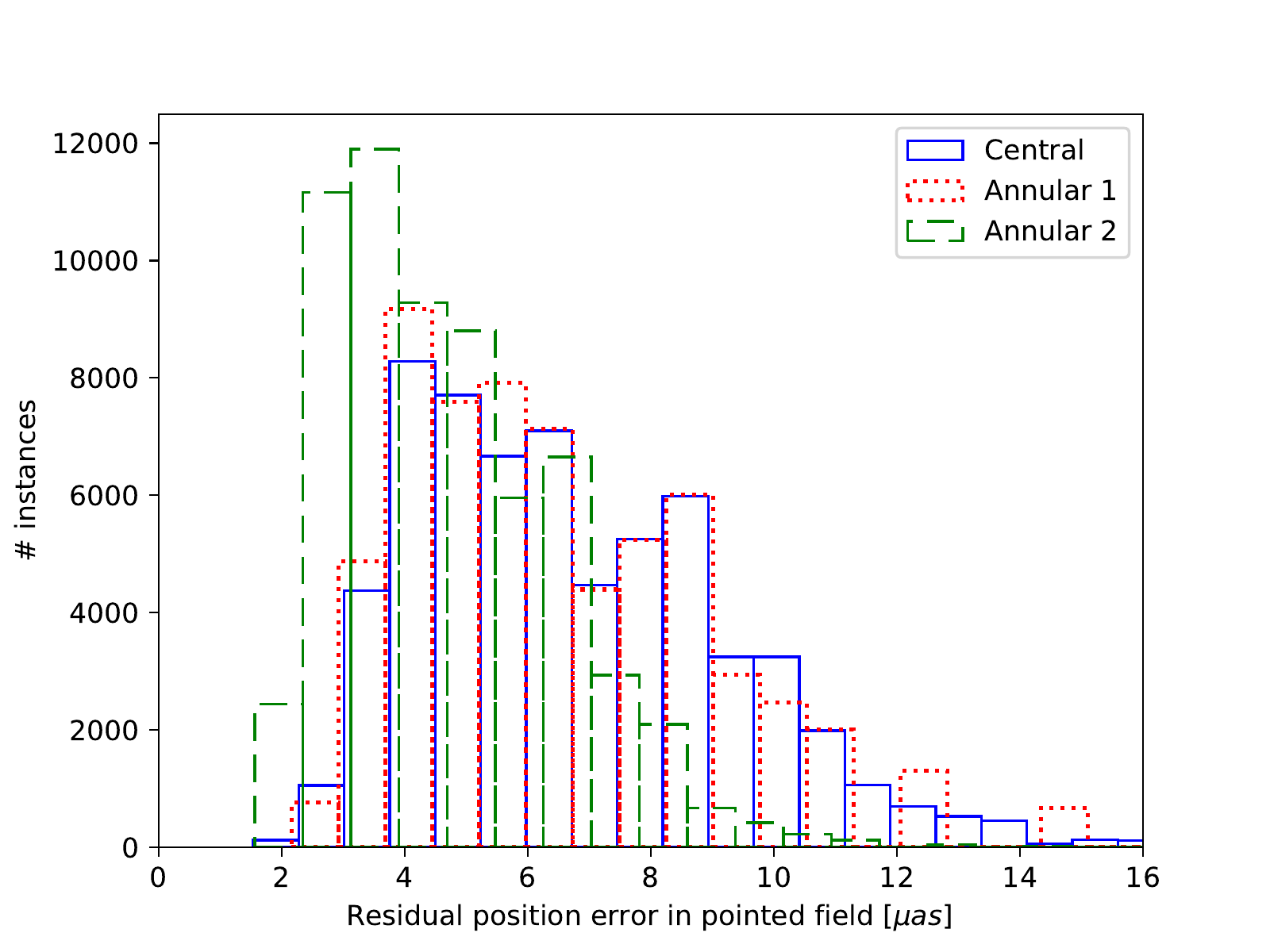}
	\includegraphics[width=0.48\textwidth,height=0.26\textheight]{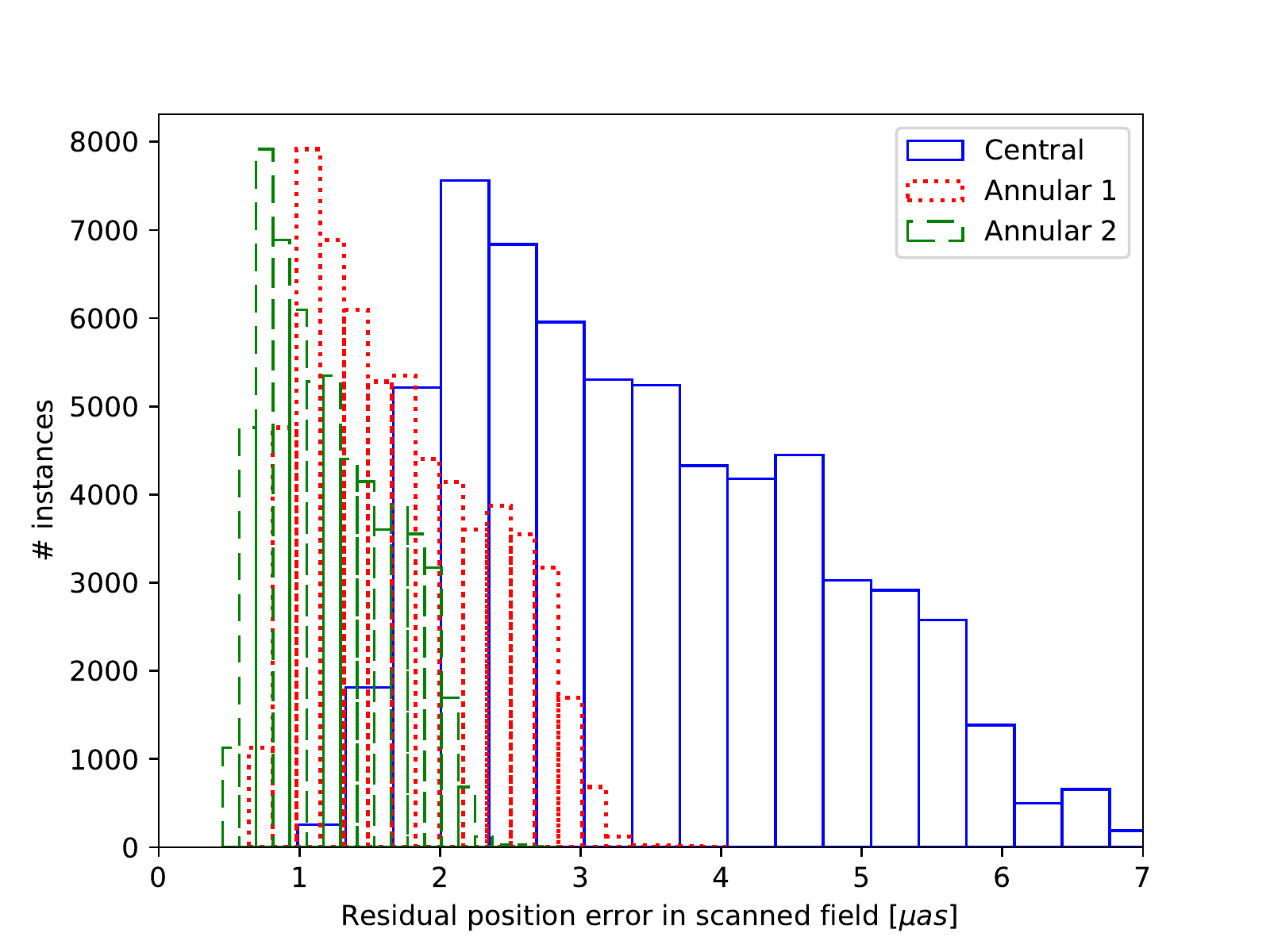}
    \caption{ Catalogue limited position errors, single pointing (left) and with 
     field scan strategy (right). 
    \label{fig:CatLimPos}}
\end{figure}

\subsection{ Catalogue residual errors }
\label{Sec:CatError}
Repeated observations of our $G \le 8\, mag$ targets, to determine their dynamics, 
is affected by residual catalogue errors from the set of field stars used as 
references, according to the preliminary remarks in Sec.\ \ref{Sec:RelAstrom}. 
Such residual errors depend therefore on observing strategy, as suggested in 
Sec.\ \ref{Sec:FieldStars}: using different field stars will reduce the random 
component. 
The selected field star populations are those shown in Fig.\,\ref{fig:FieldStars}, 
respectively for fixed pointing (left), and for a set of 15 different pointing 
offsets (right). 

\begin{figure}[ht!]
	\centering 
    \includegraphics[width=0.48\textwidth,height=0.26\textheight]{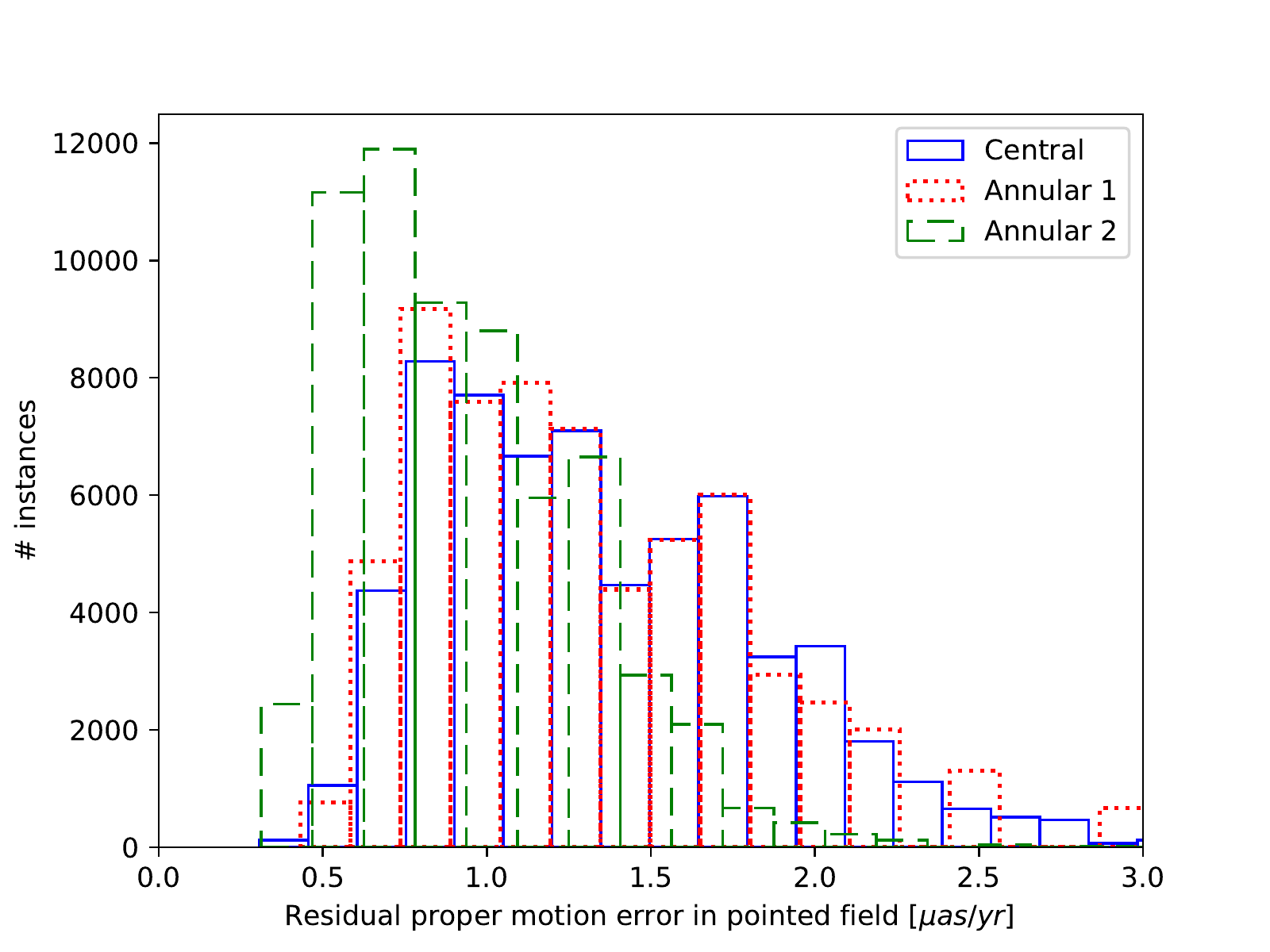}
	\includegraphics[width=0.48\textwidth,height=0.26\textheight]{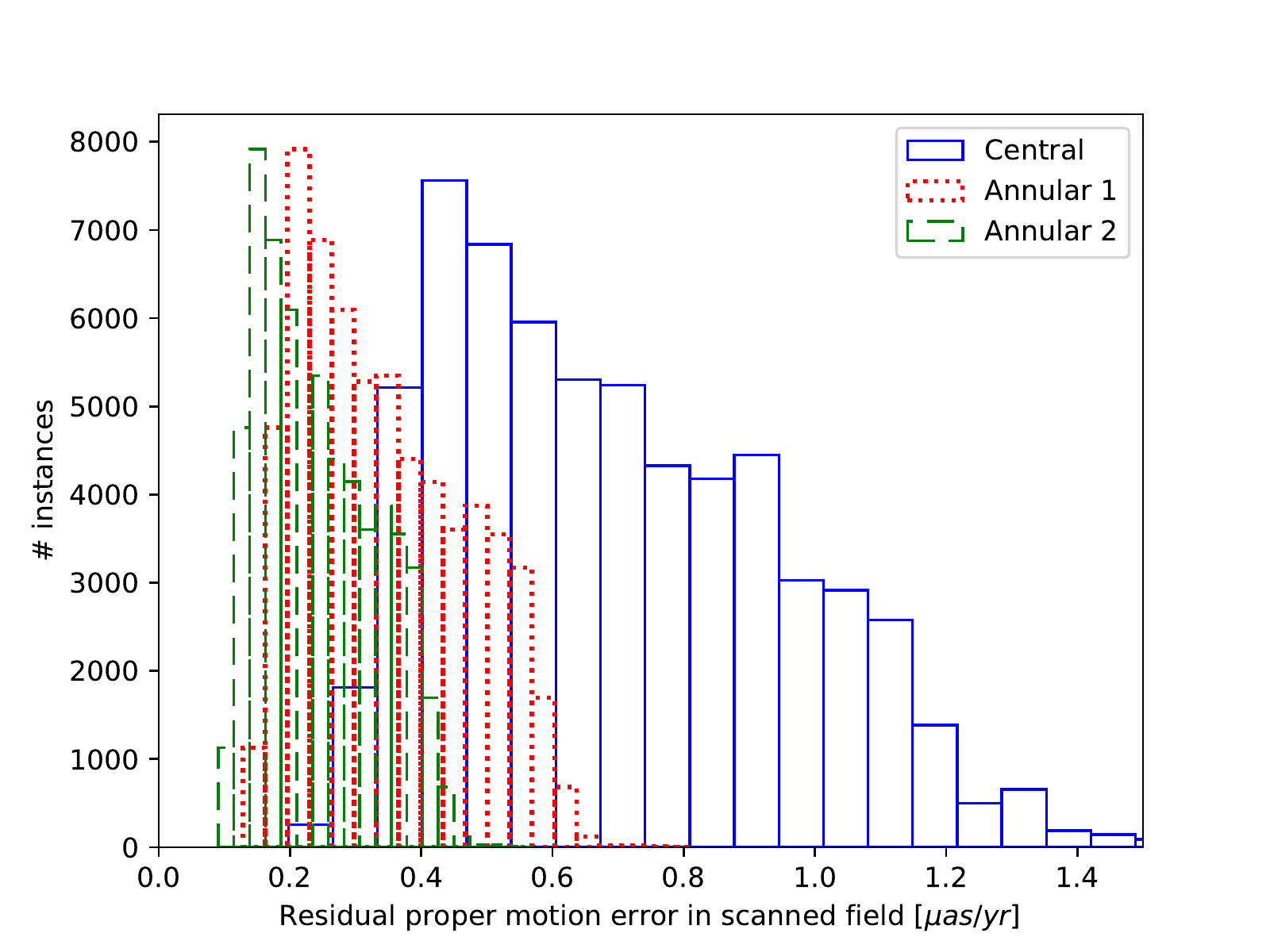}
    \caption{ Catalogue limited proper motion errors, single pointing (left) and 
    	with field scan strategy (right).\label{fig:CatLimPM}}
\end{figure}

\begin{figure}[ht!]
	\centering 
    \includegraphics[width=0.48\textwidth,height=0.26\textheight]{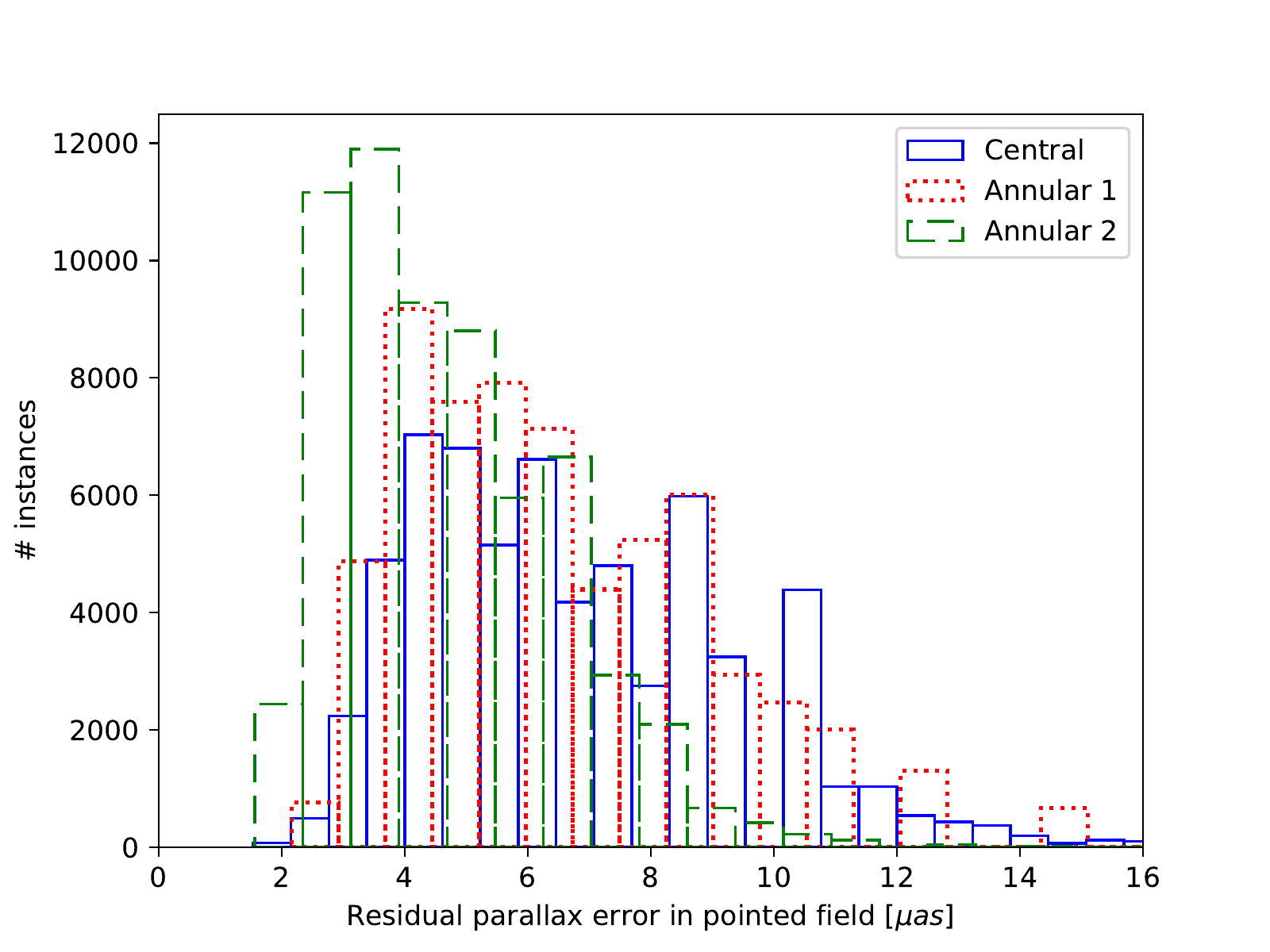}
	\includegraphics[width=0.48\textwidth,height=0.26\textheight]{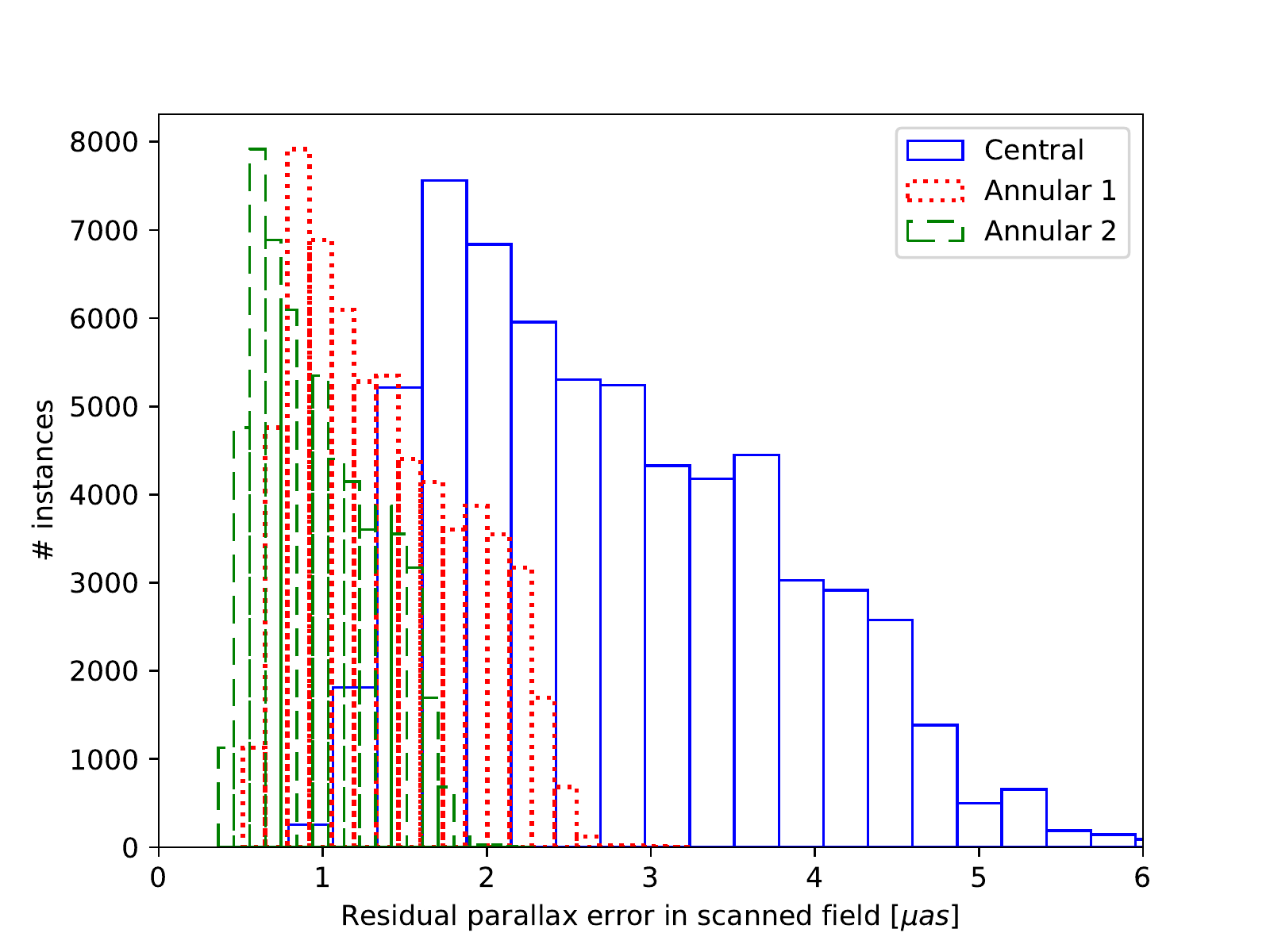}
    \caption{ Catalogue limited parallax errors, single pointing (left) and 
    	with field scan strategy (right).\label{fig:CatLimPar}}
\end{figure}

The statistics from our simulation is shown in 
Figs.\,\ref{fig:CatLimPos}, \ref{fig:CatLimPM} and \ref{fig:CatLimPar}, 
respectively for position, proper motion and parallax. 
In the left panel, the results of repeated observation of the same field, 
i.e.\ using the same reference stars for each target, are shown. 
In the right panel, the values are related to the alternative field scan  
strategy, which selects different sets of field stars within the accessible 
area around the target (15 different pointing offsets throughout visits). 
In the latter case, the larger accessible sky area provided by the 
annular field makes available many more stars than the central field 
case, with a further improvement with the width of the annular field. 
Since each star is supposed to be affected by a random catalogue error, 
the net effect is a more effective reduction of the overall uncertainty on 
the actual target position. 
Systematic catalogue errors are supposed to be negligible. 

\begin{table}
    \centering
    \caption{Residual catalogue error 
    \label{tab:ResCat}}
    \vspace{2mm}
    \renewcommand{\arraystretch}{1.2} 
\begin{tabular}{lccc}
 & Central  & Annular 1 & Annular 2\tabularnewline
\hline 
Median position, BP pointing {[}$\mu as${]} & 6.25 & 6.06 & 4.41 \tabularnewline
RMS position, BP pointing {[}$\mu as${]} & 2.59 & 2.57 & 1.76 \tabularnewline
Median position, scan {[}$\mu as${]} & 3.25 & 1.64 & 1.16 \tabularnewline
RMS position, scan {[}$\mu as${]} & 1.30 & 0.62 & 0.44 \tabularnewline
Median proper motion, BP pointing {[}$\mu as/yr${]} & 1.25 & 1.21 & 0.88 \tabularnewline
RMS proper motion, BP pointing {[}$\mu as/yr${]} & 0.52 & 0.51 & 0.35 \tabularnewline
Median proper motion, scan {[}$\mu as/yr${]} & 0.65 & 0.33 & 0.23 \tabularnewline
RMS proper motion, scan {[}$\mu as/yr${]} & 0.26 & 0.12 & 0.09 \tabularnewline
Median parallax, BP pointing {[}$\mu as${]} & 6.25 & 6.07 & 4.41 \tabularnewline
RMS parallax, BP pointing {[}$\mu as${]} & 2.58 & 2.57 & 1.75 \tabularnewline
Median parallax, scan {[}$\mu as${]} & 2.60 & 1.31 & 0.93 \tabularnewline
RMS parallax, scan {[}$\mu as${]} & 1.04 & 0.49 & 0.35 \tabularnewline
\end{tabular}
\end{table} 

The results are in the range of a few $\mu as$ on positions and parallaxes, and 
$\sim 1 \mu as/yr$ on proper motion, just using a total observing time of 15 hours 
per target. 
This may be adequate to some science topic, but insufficient for others; however, 
{\em smart tuning of the observing strategy appears able to provide impressive 
performance even with parsimonious usage of limited time resources}. 
Relevant values are summarised in Table\ \ref{tab:ResCat}.

\subsection{ Catalogue maintenance and improvement }
\label{Sec:CatImpro}
The astrometric measurements from observations may be used to improve, up to 
a point and in suitable conditions, the catalogue precision on field stars, 
or at least to preserve it against natural degradation. 
Not only field star errors can be averaged out effectively to get proper determination 
of any target's kynematic parameters, but in principle astrometry on field stars 
can be improved as well, assuming they are observed repeatedly throughout the 
mission lifetime. 
We again assume three observations per year over five years, uniformly distributed. 
This simulation is performed on the whole sample of Gaia objects down to 
$G = 12 \, mag$, each considered as a single star with a simple linear motion 
described by the five astrometric parameters in the catalogue. 
Each star is located against the reference system materialised by the set of 
remaining objects, using the single epoch precision in Fig.\,\ref{fig:FieldPrec} 
(right). 

The full sample provides a realistic distribution of positions on the sky, proper 
motion and parallax, suited to our immediate goal of evaluating the potential 
improvement on astrometric parameters of field stars achievable with the 
proposed observing strategy. 
In practice, depending on target and sampling scheme selection from the main 
science case, the number of field star subject to astrometric bootstrap 
may range, e.g. between few ten thousand to few hundred thousand objects. 

\begin{figure}
	\centering 
	    \begin{minipage}[b]{0.3\textwidth}
    \includegraphics[width=\textwidth,height=0.25\textheight]{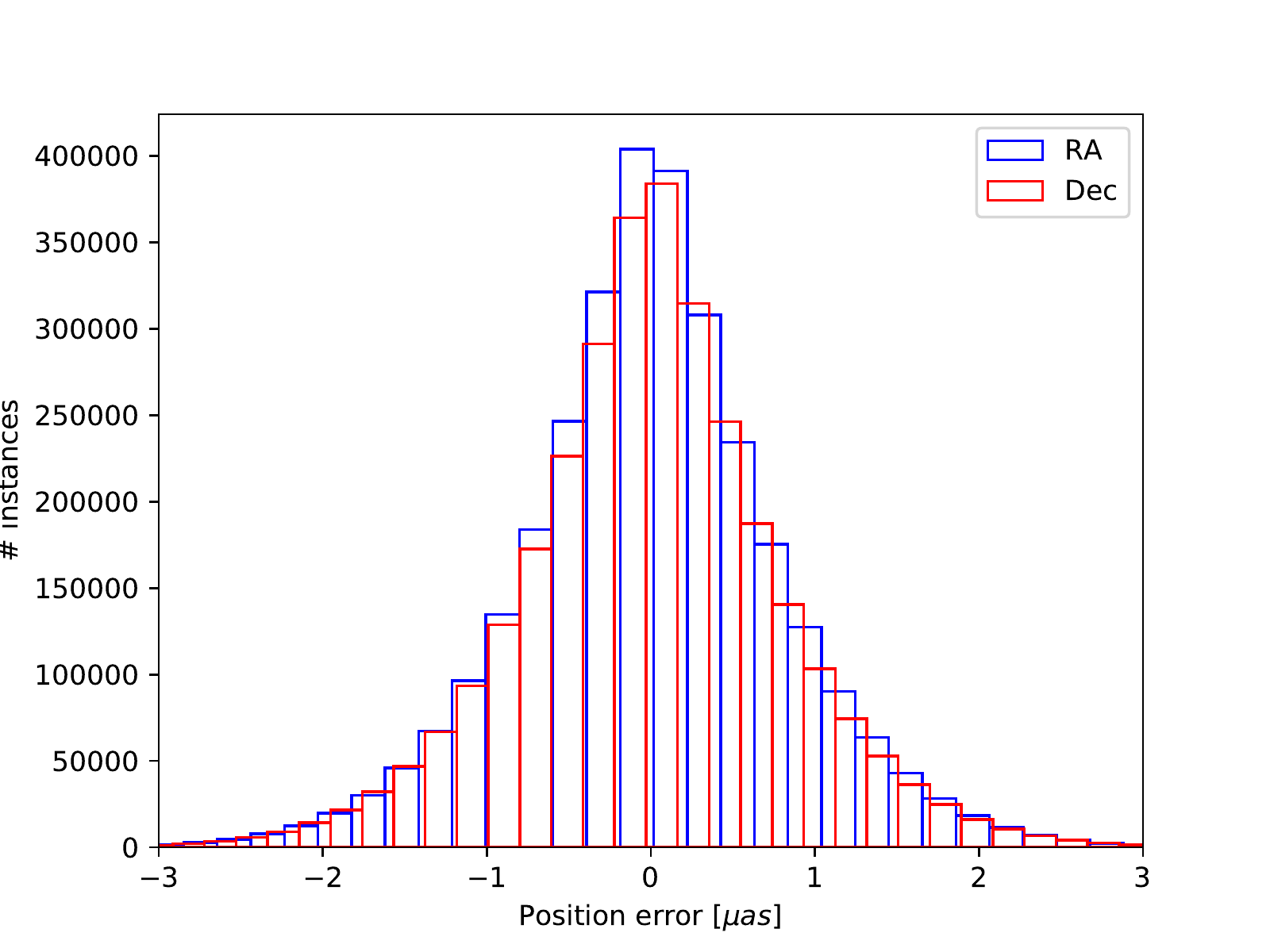}
    \caption{ Photon limited errors on reference star positions.\label{fig:CatRecPos}}
\end{minipage}\qquad
    \begin{minipage}[b]{0.3\textwidth}
    \includegraphics[width=\textwidth,height=0.25\textheight]{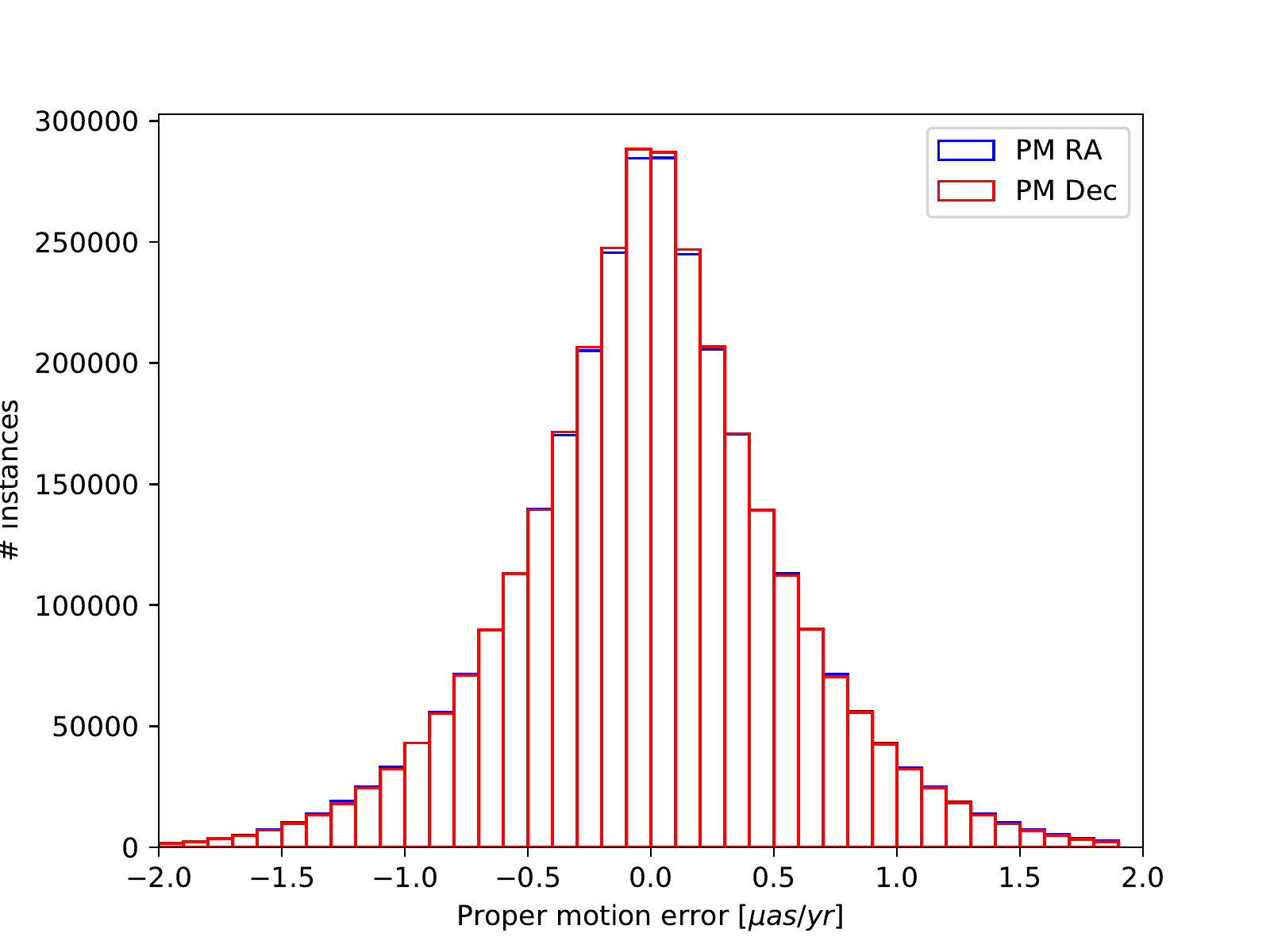}
    \caption{ Photon limited errors on reference star proper motions. 
    \label{fig:CatRecPMot}}
\end{minipage}\qquad
    \begin{minipage}[b]{0.3\textwidth}
    \includegraphics[width=\textwidth,height=0.25\textheight]{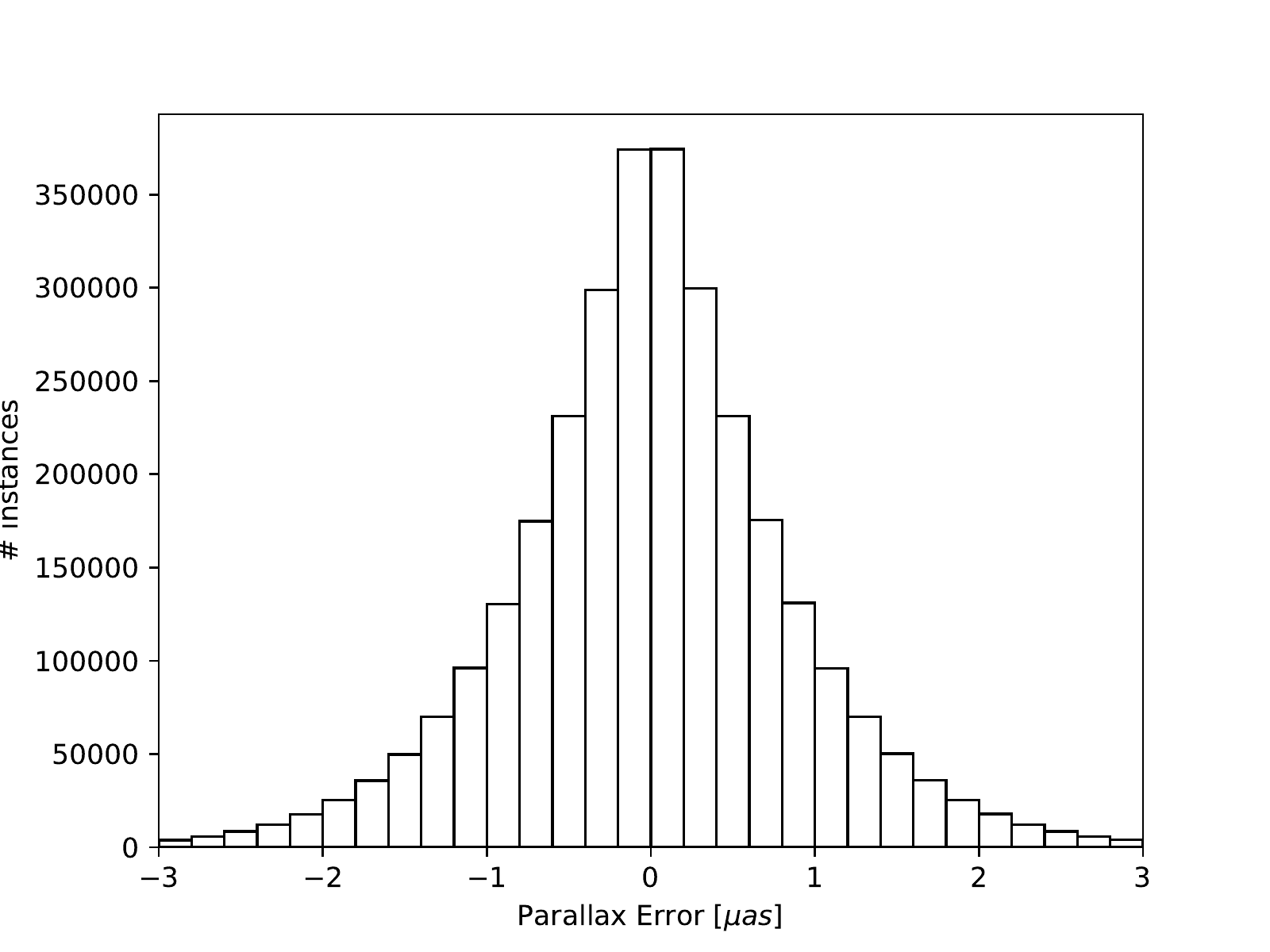}
    \caption{ Photon limited errors on reference star parallaxes. 
    \label{fig:CatRecPar}}
\end{minipage}
\end{figure}

The results of the simulation are shown in Figs.\ \ref{fig:CatRecPos}, 
\ref{fig:CatRecPMot} and \ref{fig:CatRecPar}, 
respectively for positions, proper motions and parallaxes. 
The corresponding statistics (median errors and their RMS spread) is reported in 
Table\ \ref{tab:CatImpro}; median errors improve from the central, through the thin 
annular, to the thick annular field case. 
The RMS of experimental errors may be considered as an expectation on 
the variability expected among different fields, depending mainly on 
source density. 
The estimation does not include the residual catalogue (absolute) 
errors, i.e. it is strictly {\em relative} astrometry.

\begin{table*}
    \centering
    \caption{Astrometric catalogue bootstrap by observation of field star\label{tab:CatImpro}}
    \renewcommand{\arraystretch}{1.2} 
    \begin{tabular}{lccc}
     & Central  & Annular 1 & Annular 2\tabularnewline
    \hline 
    Median position error {[}$\mu as${]}            &  3.25 & 1.64 & 1.16  \\
    RMS position error {[}$\mu as${]}               &  1.30 & 0.62 & 0.44  \\
    Median proper motion error {[}$\mu as/yr${]}    &  0.65 & 0.33 & 0.23  \\
    RMS proper motion error {[}$\mu as/yr${]}       &  0.26 & 0.12 & 0.09  \\
    Median parallax error {[}$\mu as${]}            &  2.60 & 1.31 & 0.93 \\
    RMS parallax error {[}$\mu as${]}               &  1.04 & 0.49 & 0.35 \\
    \end{tabular}
\end{table*}

\subsection{ Pushing toward fainter magnitude }
\label{Sec:DeeperMag}
Most of the evaluations in previous sections have been performed on a slice of 
the Gaia EDR3 catalogue limited to $G \le 12\, mag$, for reasons of practicality. 
The number of available field stars obviously increases by setting 
a fainter limiting magnitude. 
It may be noted that fainter stars, e.g.\ down to $G \le 15\, mag$, 
are still quite bright for a $1\,m$ class telescope, requiring elementary 
exposure time below $1\,min$, and providing  
an epoch location uncertainty of $\lesssim 10\,\mu as$ in one hour. 
It may be expected that, in general, random error components on the reference frame 
materialization (implemented by the whole set of field stars) will improve by setting 
a fainter limiting magnitude, i.e. using more stars, albeit each affected by 
comparably large individual errors. 

\begin{table*}[h]
	\centering
	\caption{ Number of field stars as a function of magnitude \label{tab:LimMag}}
	\renewcommand{\arraystretch}{1.2} 
	\begin{tabular}{lcccccc}
		G  & \multicolumn{2}{c}{ Central } & \multicolumn{2}{c}{ Thin Annular } & 
		\multicolumn{2}{c}{Thick Annular} \\
		mag & Median   & RMS & Median   & RMS & Median   & RMS \\
		\hline 
		12 &  59.0 &    70.9 &   185.5 &   220.6 &   371.0 &    441.1 \\
		13 & 124.0 &   134.9 &   400.8 &   659.7 &   801.5 &  1,319.4 \\
		14 & 258.5 &   458.9 &   880.9 & 2,082,4 & 1,761.8 &  4,164.8 \\
		15 & 483.0 & 1,491.3 & 1,903.6 & 6,404.5 & 3,807.2 & 12,808.9 \\
	\end{tabular}
\end{table*} 

\begin{figure}
	\centering 
	\includegraphics[width=0.48\textwidth,height=0.26\textheight]{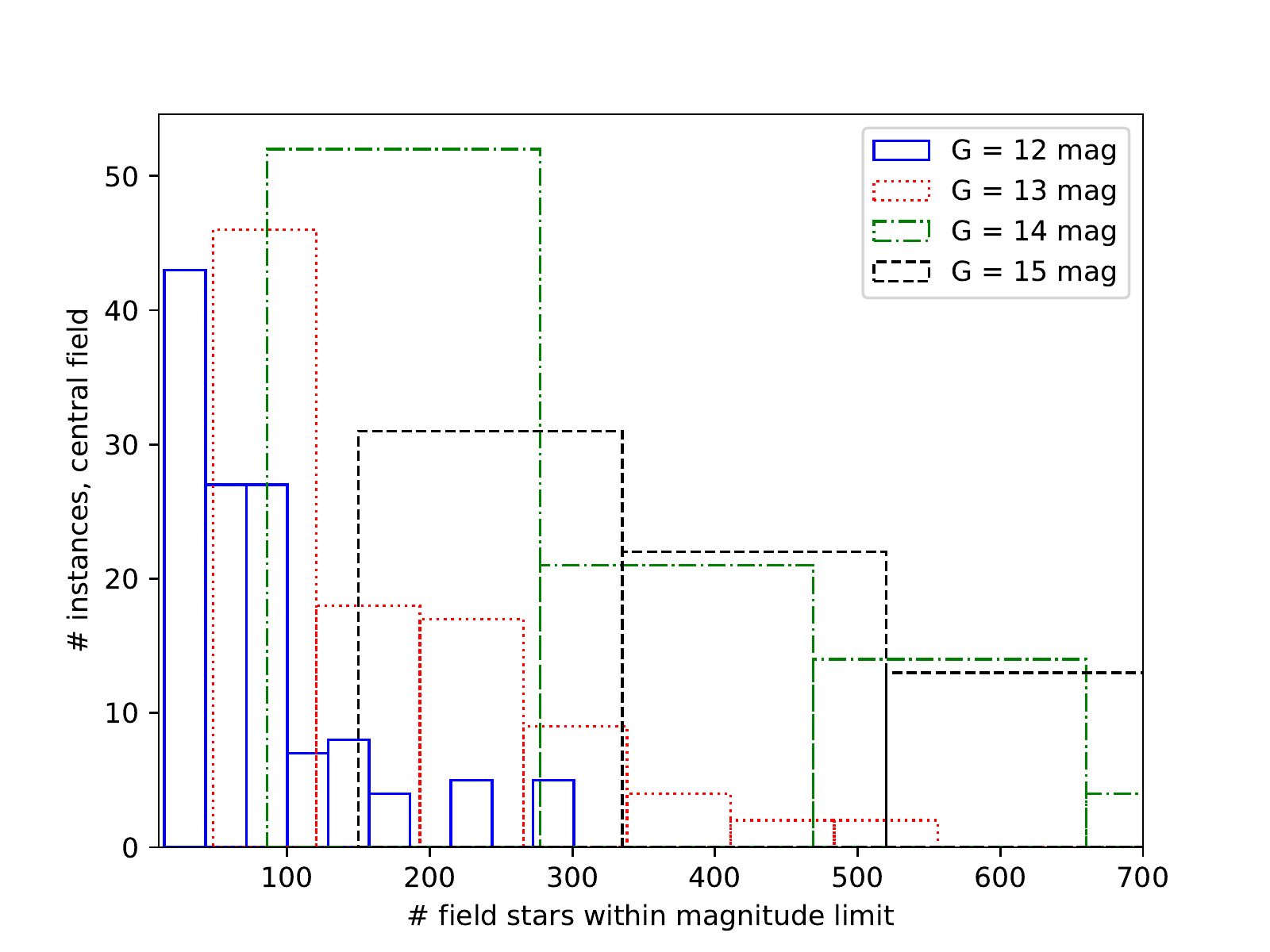}
	\includegraphics[width=0.48\textwidth,height=0.26\textheight]{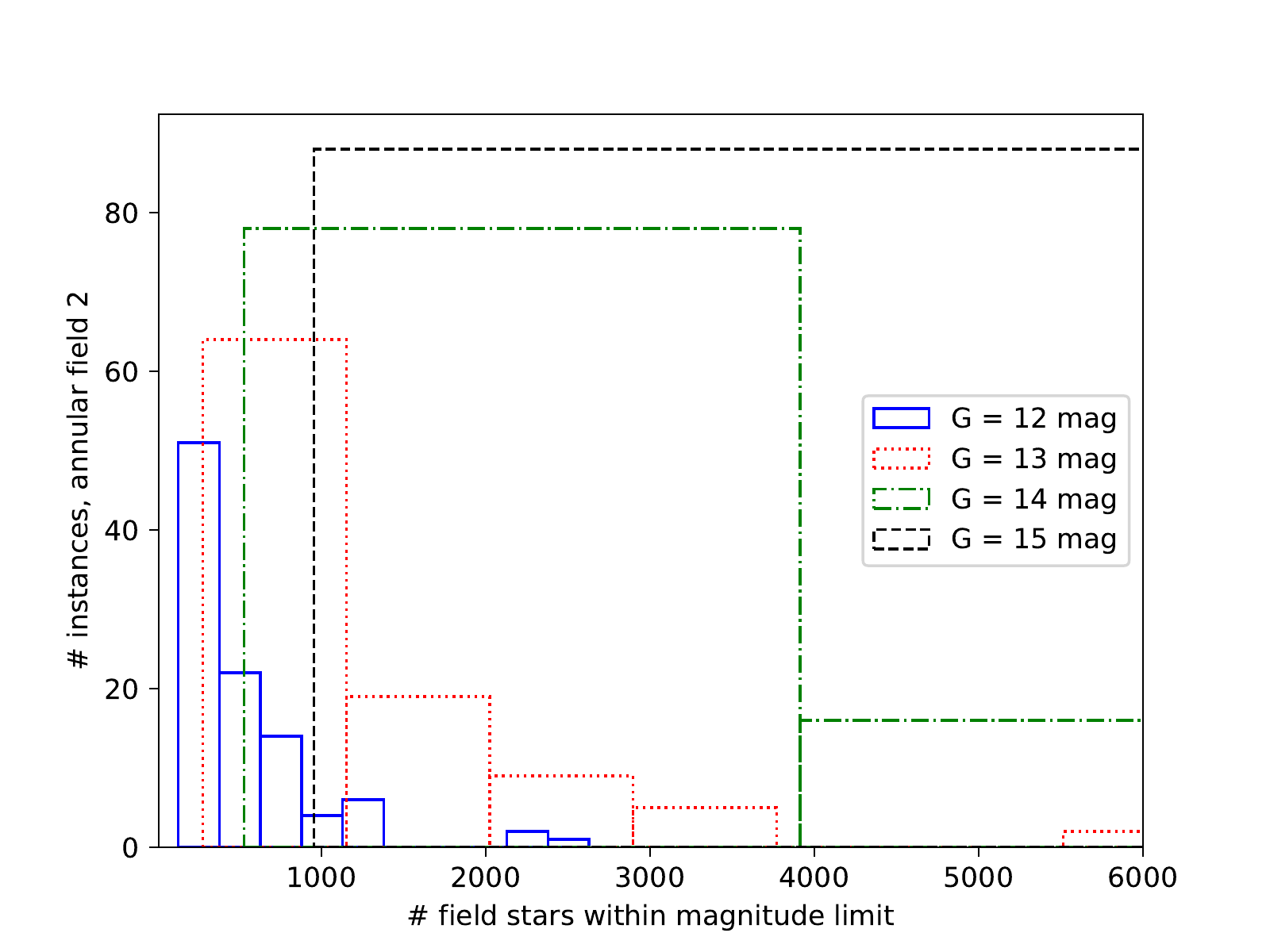}
	\caption{ Number of field stars as a function of the limiting magnitude, 
		for the central (left) and thick annular field (right).\label{fig:NStarMag}}
\end{figure}

The star density in the range $12 \le G \le 15\, mag$ is more dependent 
on Galactic latitude, compared to brighter sources, which are somewhat 
more uniformly distributed. 
A simple assessment of average sky density has been performed by selecting 
from the Gaia catalogue a sample of all sources over the range $G \le 12\, mag$, 
within a $2^\circ$ radius around 100 bright objects in the range 
$4.2 \lesssim G \lesssim 6\,mag$, 
distributed on rather sparse positions on the sky. 
\\ 
The number of stars down to a given magnitude in this sample 
is shown in Fig.\,\ref{fig:NStarMag}, respectively for the central (left) 
and thick annular (right) case. 
The thin annular case, intermediate between the others, is omitted for the sake 
of brevity. 
The typical number of reference stars increases with the limiting magnitude, 
and it is significantly larger in the latter case, as expected. 
Statistical values are listed in Table\ \ref{tab:LimMag}; the large 
spread evidenced by the RMS is mainly due to the variation in star population 
with Galactic latitude.

\section{ Conclusions }
\label{Sec:conclusions}
The potential for relative astrometry of a $1\,m$ class space telescope 
endowed with an annular field of view with $\sim 1^\circ$ radius is 
evaluated by comparison with a conventional instrument concept with 
a compact focal plane detector. 
Basic concepts of astrometry as a tool for crucial parameter determination 
are recalled, in particular with respect to real mass determination of 
exo-planetary systems. 
We evidence that smart observing plans, exploiting prior knowledge on 
individual targets, result in efficient measurement, e.g.\ pinpointing 
orbit inclinations, and therefore true planet masses, with a minimum number 
of dedicated observations. 
The statistics on reference field stars brighter than $G = 12\,mag$, for 
targets down to $G = 8\,mag$, is evaluated on the Gaia EDR3 catalogue; 
the potential of sufficiently wide sets of Gaia sources as local 
materialization of the reference frame at the $\mu as$ level is 
discussed. 
The annular field provides the capability of finding adequate reference 
objects to a larger distance from the selected target than the central field, 
thus improving 
on the overall astrometric measurement because more and/or brighter 
field stars are made available. 
In particular, in our scenario, each target may find in the annular field 
individual reference stars typically $2\,mag$ brighter than in the central 
field, or many more stars (four to seven times as many in the example 
described), resulting in a more reliable reference frame. 
Repeated observation of selected sources (several ten thousand, to few 
hundred thousand) may ``refresh" their astrometric parameters, mitigating 
the natural catalogue degradation with time. 

\section*{ Acknowledgments } 
The activity has been partially funded by a grant from the Italian Ministry of 
Foreign Affairs and International Cooperation (ASTRA: Astrometric Science and 
Technology Roadmap), and by the Italian Space Agency 
(ASI) under contract 2018-24-HH.0. 

\bibliographystyle{aasjournal}

\end{document}